\documentclass[nolinenumbers]{aastex631}
\usepackage{hhline,colortbl}
\usepackage{amsmath}
\usepackage{placeins}

\usepackage{svg}
\usepackage{amssymb}
\begin{document}

\newcommand{\xmm}{{\small {\it XMM-Newton}}} 
\newcommand{\xrgs}{{\small {\it XMM-Newton}/RGS}} 
\newcommand{\xpn}{{\small {\it XMM-Newton}/EPIC-PN}} 
\newcommand{\rgs}{{\small RGS}}
\newcommand{\letg}{{\small LETG}}

\newcommand{\chandra}{{\small {\it Chandra}}}
\newcommand{\chletg}{{\small {\it Chandra}/LETG}}
\newcommand{\hetg}{{\small {\it Chandra}/HETG}}
\newcommand{\xrism}{{\small{\it XRISM}}} 
\newcommand{\resolve}{{\small {\it Resolve}}}
\newcommand{\xrt}{{\small {\it Swift}/XRT}} 
\newcommand{\swift}{{\small {\it Swift}}} 
\newcommand{\nicer}{{\small {\it NICER}}} 

\newcommand{\gro}{GRO\,J1655-40}
\newcommand{\yzret}{{\small YZ\,Ret}}

\newcommand{\blue}[1]{{\textcolor{blue}{#1}}}
\newcommand{\red}[1]{{\textcolor{red}{#1}}}
\newcommand{\green}[1]{{\textcolor{green}{#1}}}

\title
{Mixing of hot shocked plasma with cold gas in Nova YZ\,Ret 2020}

\correspondingauthor{Sharon Mitrani}
\email{sharonm@campus.technion.ac.il}

\author{Sharon Mitrani}
\affiliation{Department of Physics, Technion, Haifa 32000, Israel}

\author{Ehud Behar}
\affiliation{Department of Physics, Technion, Haifa 32000, Israel}
\affiliation{MIT Kavli Institute for Astrophysics and Space Research, Massachusetts Institute of Technology, Cambridge, MA 02139, USA}

\author{Jeremy J. Drake}
\affiliation{Lockheed Martin, 3251 Hanover Street, Palo Alto, CA 94304}

\author{Marina Orio}
\affiliation{Department of Astronomy, University of Wisconsin, 475 N. Charter Str., Madison, WI 53706, USA}
\affiliation{INAF-Padova, vicolo Osservatorio 5, I-35122 Padova, Italy}

\author{Kim Page}
\affiliation{School of Physics and Astronomy, The University of Leicester, University Road, Leicester LE1 7RH, UK}

\author{Valentina Canton}
\affiliation{INAF-Padova, vicolo Osservatorio 5, I-35122 Padova, Italy}

\author{Jan-Uwe Ness}
\affiliation{European Space Agency (ESA), European Space Astronomy Centre (ESAC), Camino Bajo del Castillo s/n, E-28692 Villanueva de la Cañada, Madrid, Spain}

\author{Kirill Sokolovsky}
\affiliation{Department of Astronomy, University of Illinois at Urbana-Champaign,
1002 W. Green Street, Urbana, IL 61801, USA}
\affiliation{Sternberg Astronomical Institute, Moscow State University, Universitetskii
pr. 13, 119992 Moscow, Russia}


\begin{abstract}
The origin of bright X-ray emission lines that appear late in a nova eruption remains largely a puzzle. 
We present two high-resolution X-ray grating spectra of the classical nova YZ Ret, observed 77 and 115 days post-eruption, using \xmm\ and \chandra , respectively. Both spectra feature resolved emission lines blueshifted by $v = -1500$\,km\,s$^{-1}$ and broadened by $\sigma_v=500$\,km\,s$^{-1}$.
%If the outflow had such a velocity over 38 days, the plasma propagated a distance of $5\times10^{14}$\,cm. %with little change to the ionization state. %which rules out a photo-ionized plasma origin. 
The two spectra are well described by a collisionally ionized plasma of $kT\sim 70$\,eV that dimmed by a factor of $\sim40$ %over the 38 day period 
between the two exposures. 
The spectra also show narrow radiative recombination continua (RRCs) of C$^{+4}$, C$^{+5}$, and N$^{+5}$, indicating the interaction of the hot ionized plasma with cold electrons of $kT\sim 2$\,eV. 
%Using the high intensity of 
The high-$n$ Rydberg series of C$^{+4}$ is anomalously bright, allowing us to measure the electron density through continuum lowering, which is in  agreement
 with the He-like N$^{+5}$ density diagnostic of
 $n_e=(1.7\pm0.4)\times10^{11}$\,cm$^{-3}$.
The high population of these high-$n$ levels constitutes the best evidence to date of charge exchange (CX) with neutral H in an astrophysical ionized plasma.
The remarkable fact that the velocity and plasma temperature are the same after 38\,days, despite the high density and decreasing flux is evidence for ongoing heating. % which we ascribe to a \
We suggest the heating is due to a reverse shock in the nova ejecta, which forms a thin X-ray shell. %n outgoing expanding shock. 
The narrow RRCs and CX are attributed to direct mixing %of the shocked hot medium 
with cold gas, which overtakes the hot plasma either from the shock front, or 
through the contact discontinuity.

%Density diagnostics using N$^{+5}$ lines and continuum lowering of C$^{+4}$ of $n_e=(1.7\pm0.4)\times10^{11}$\,cm$^{-3}$ during the first epoch are consistent with the expanding shell scenario and the decrease of X-ray flux and density.

\end{abstract}


%%%%%%%%%%%%%%%%%%%\keywords{Classical Novae (251) --- Ultraviolet astronomy(1736) --- History of astronomy(1868) --- Interdisciplinary astronomy(804)}


\section{Introduction} \label{sec:intro}
Classical and recurrent novae are interacting binary systems hosting
a white dwarf (WD). Their eruptions are attributed to a
thermonuclear runaway (TNR) in the atmosphere of the WD,
which has been accreting material from its binary companion. Models predict that TNR is followed by a radiation
driven wind, which depletes the
accreted envelope \citep{Starrfield2012, Wolf2013}, or
by a wind originating in the common envelope, most likely because of
double Roche-lobe filling \citep[see][]{Shen2022}.

Novae eruptions are luminous at all wavelengths, from $\gamma$-rays to
radio. X-rays specifically provide an important window
to understand their physics. Early in the nova eruption,
the X-rays are attributed to powerful shocks in the outflow. Line-resolved X-ray grating spectra have been successfully described by collisional ionization steady-state models \citep[see
the discussions in][]{Orio2012, Drake2016, Peretz2016, Orio2020,
 Chomiuk2021}. 
 However, transient states occur early in the eruption \citep{Orio2023, Islam2023}. 
 %In most novae, the X-ray luminosity in the 0.2–10\,keV range peaks at 10$^{34}$\,erg\,s$^{-1}$. 
%Symbiotic novae can be a factor of 100 more luminous, a
%fact that has been attributed to the collision of the nova ejecta
%with the red giant wind. 
Nova shocks are
powerful enough that they can produce $\gamma$-ray emission
\citep{Franckowiak2018}, of either leptonic
(e.g., inverse Compton off of accelerated electrons) or hadronic origin (proton interactions producing neutral pions that emit $\gamma$-rays as they decay).
 
After a few days to almost a year, most novae transition to the supersoft phase, where the X-ray luminosity can exceed 10$^{38}$ erg s$^{-1}$ (Eddington level for $\simeq$ 1 M$_\odot$), with all the flux emitted  below 0.8\,keV \citep[see review by][]{Orio2012}. 
This happens because the WD atmosphere contracts at constant bolometric luminosity, so that the effective temperature keeps on increasing and the peak wavelength of the emission moves from the optical range, to the UV and extreme UV, and eventually to the soft X-rays within weeks.
The source continues to be powered by shell burning. 
The central source appears as a {\it thermal} continuum X-ray supersoft source (SSS), with a peak temperature up to $10^6$\,K, for a time lasting from days to a few years \citep{Yaron2005}.
The more massive the WD, the shorter is the SSS phase \citep{Yaron2005, Starrfield2012, Wolf2013}.

 The 2020 nova \yzret\ discussed in this paper appeared to follow the trend of brightening and softening, as expected in the emergence of the SSS, however the X-ray luminosity remained
much below the level of other novae \citep[e.g.][]{Orio2022}. 
When the high resolution X-ray observations were triggered, quite surprisingly only an emission line spectrum was observed, with no sign of the WD continuum (SSS) \citep{Sokolovsky2022}.
The SSS could be obscured by an accretion disk, %of material around it, which 
indicating a high inclination system \citep{Ness2013}. In this paper, we analyze and discuss the puzzling X-ray spectra of the nova, at X-ray maximum flux and during the decay phase.

\subsection{Nova outburst of YZ Ret}
Nova YZ Ret nova erupted in a known binary with a WD. For a summary of previous observations, see \citet{Sokolovsky2022}. It was discovered in outburst at 5th magnitude on 2020
 July 15, and the optical maximum was later
 discovered to have occurred on 2020 July 11 at 3.7\,mag \citep{Kaufman2020}.
 % \citep{Kaufman2020}.
Hence, \yzret\ became visible with the naked eye and the second-brightest nova in a decade.  
Optical spectroscopic follow-up confirmed it to be a nova \citep{Aydi2020}. 
%Early spectra were dominated by a blue continuum with superimposed broad high-order Balmer
%absorption lines and weak H$\beta$ absorption
%(probably filled with emission), resembling dwarf-novae and nova-like objects in outburst 
 %but later spectroscopic observations measured a typical nova spectrum.
\citet{Kaufman2020} described
the spectrum
 after the onset of the eruption as that of a Fe II-type nova
 \citep[according to the classification scheme of][]{Williams1992},
but \citet{Carr2020} reported a He/N-type on
 later days. 
 There is now a common understanding that these are
 successive phases in the spectral evolution of novae \citep{Aydi2023}.
 From an overabundance of oxygen and prominent UV emission lines in the UV, \citet[][]{Izzo2020} concluded that the nova erupted on an ONe WD. 
 However, we find no oxygen in the X-ray spectrum.

An X-ray flash detected from \yzret\ immediately after the outburst was ascribed to the fireball phase, namely an early SSS phase of a few hours, when the nova is still undergoing the TNR %, and before it is visible in the optical 
\citep{Konig2022}.
\yzret\ was detected in $\gamma$-rays (0.1-1\,GeV) with {\sl Fermi-LAT}, and in hard X-rays (3-79\,keV) with {\sl NuSTAR}, during its first two weeks.
\citet{Sokolovsky2022} 
fitted the X-ray broad band spectra
with unusual abundances from their model, namely a collisionally-ionized plasma either deficient in iron or overabundant in oxygen and nitrogen.
They also showed the \xmm\ high resolution grating spectrum, which is the focus of the present paper, but did not analyze it.

YZ Ret was classified as a VY Scl system %because the Sloan Survey 
%indicated a mean magnitude V=16.6 with two episodic “states” at magnitude
%V=18-19
\citep[][and references therein]{Sokolovsky2022}.
VY Scl binaries were defined by \citet{Honeycutt2004}
 as cataclysmic variables undergoing fading of the optical light
by 1.5 to 7 mag over less than 150 days, and these states recur over timescales from weeks to years.
In the more common high state, 
VY Scl binaries have large optical and UV luminosity. This
is interpreted as evidence that most of the time mass
transfer onto the WD occurs at a rate $\dot m > 10^{-10}$
M$_{\odot}$ yr$^{-1}$, necessary to sustain an accretion disc in a
stable hot state in which dwarf novae outbursts are
suppressed. The low states have been attributed to
a sudden drop of $\dot m$ from the secondary, or even to
a total cessation of mass transfer \citep{King1998, Hessman2000},
 perhaps due to spots on the
surface of the secondary covering the L1 point \citep{Livio1994}
 or non-equilibrium of the irradiated
atmosphere of the donor \citep{Wu1995}. Monitoring the X-ray emission of
a few of these peculiar binaries, \citet{Zemko2014}
ruled out 
non-explosive thermonuclear burning.
%The outburst of 
%YZ Ret has  proven that even if thermonuclear
%burning does occur, it can end in a nova outburst,
%instead of being steady and without outflows as expected for the steady burning in some putative SNe Ia progenitors.

 The observations included in this article were triggered after {\sl Swift} XRT monitoring showed the emergence of a soft and luminous object, with the aim of studying the WD emission in the SSS phase. 
 It was soon realized from \xmm\ \citep{Sokolovsky2022} and {\sl NICER} spectra \citep{Orio2022} that the X-rays from \yzret\ cannot be attributed to the WD. %but rather to the ejecta and circumstellar material. 
 Detecting such soft X-ray flux %from the outflow 
 from Galactic novae is not often possible due to Galactic column densities of $N_{\rm H}>10^{21}$\,cm$^{-2}$.
 However, \yzret\ is at a Galactic latitude of $-46^\circ$, with a 
 low column density of only $N_{\rm H} = 1.2 \times 10^{20}$\,cm$^{-2}$ \citep{Kalberla2005}. 
 The distance to the nova is fairly well determined with GAIA, as $2.5 \pm 0.3$\,kpc \citep{Bailer2021}, which allowed %, with  {\sl NICER} data, 
 to estimate its X-ray luminosity at a few $10^{35}$\,erg\,s$^{-1}$.
 Indeed, this is an order of magnitude smaller than even the least luminous SSS-WD \citep[generally, few 10$^{36}$\,erg\,s$^{-1}$ to
 few 10$^{38}$\,erg\,s$^{-1}$,][]{Starrfield2012}.

 The two high resolution observations presented here confirm X-ray line emission with no photospheric SSS whatsoever. The SSS, thought to be the smoking gun of nuclear burning, has not been observed also in a few other novae and non-outbursting WD-SSS. %\citep[see sample of][]{Ness2013}. 
 \citet{Ness2013} explained the missing SSS with a non-disrupted (or quickly rebuilt) accretion disk in systems with high inclination. 
 %In some cases the SSS fades and only a X-ray spectrum with emission lines is observed \citep{Ness2005, Nelson2008, Rohrbach2009}.
 Notwithstanding, and as we show in the following sections, the spectra of \yzret\ are like no other, providing unprecedented insights into the atomic processes of shocked novae outflows. 
 We describe the observations in Sec.\,\ref{sec:obs}, the spectral diagnostic methods in Sec.\,\ref{sec:analysis}, and their rich application to the spectra in Sec.\,\ref{sec:spec}. In Sec.\,\ref{sec:full} we report global fits to the spectra.
 We discuss all of the results in Sec.\,\ref{sec:discussion} and list our conclusions in Sec.\,\ref{sec:conc}.

\section{Observations} \label{sec:obs}
\yzret\ was monitored in outburst with the {\sl Swift} X-ray Telescope \citep[XRT,][]{Burrows2005} over 155 days with a cadence that ranged from every 2 days to weekly monitoring. 
Two X-ray gratings observations of \yzret\ were triggered 77 and 115 days after the eruption; the first observation with the \xmm\  Reflection Grating Spectrometers \citep[\rgs,][]{denHerder2001}, and the second one with the \chandra\ Low Energy Transmission Grating \citep[\letg,][]{Brinkman2000}.

\subsection{Swift XRT lightcurve}
\begin{figure}[h]
 \centering
 \includegraphics[width=0.67\textwidth]{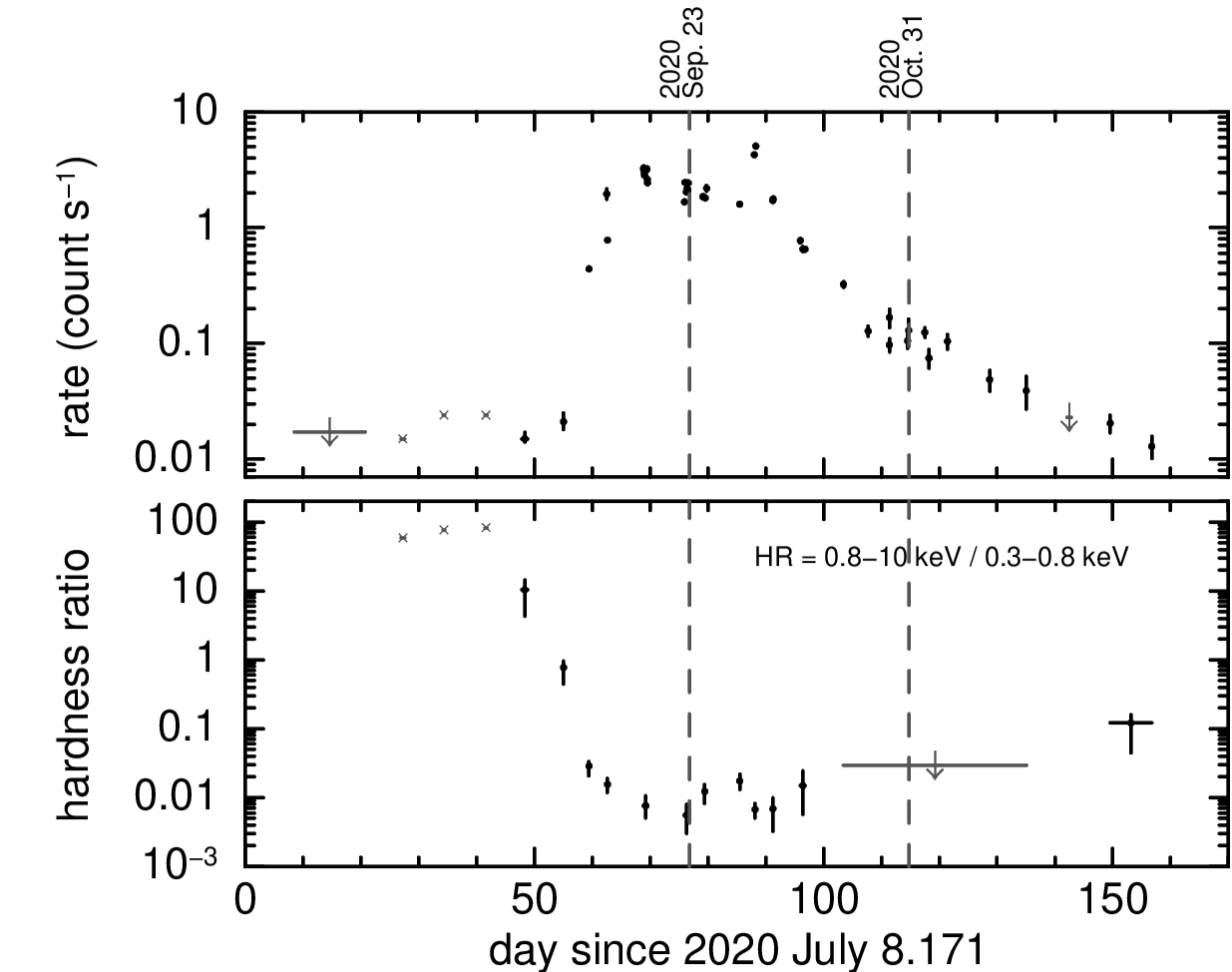}
 \caption{Swift XRT count rate (top panel) and hardness ratio (lower panel) measured in the first 155 days of the nova outburst. The vertical dashed lines indicate the observations of \xmm\ (day 77) and Chandra (day 115). The observations obtained between days 27 and 42 were affected by high
background emission below 0.5 keV. In these cases, the spectrum was fitted
over 1-10 keV, and the model extrapolated down to 0.3 keV to estimate the
count rate over the full 0.3-10 keV band. These data points are marked with x.
} 
 \label{fig:swiftlc}
 \end{figure}

Fig.\,\ref{fig:swiftlc} shows the XRT lightcurve featuring a rapid increase in count rate and a rapid decrease in hardness after day 50. 
%However, as outlined above and found with {NICER} by \citet{Orio2022}, the X-ray luminosity of \yzret\ was inconsistent with the WD being completely visible.  
At about day 90 the count rate started decreasing monotonically until day 160 when monitoring stopped. The \rgs\ observation at day 77 caught the nova close to its highest state, while the \letg\ observation was carried out when the XRT count rate was more than an order of magnitude lower. Both grating spectra show a rich emission-line spectrum with no obvious continuum. 
  
  %As Fig.\,\ref{fig:swiftlc} shows, the hardness ratio
   %calculated as the ratio of counts between 0.8 keV and 10 keV and
    %the counts between 0.3 and 0.8 keV had a rapid decrease as the
     %X-ray luminosity increased, and remained around 0.01 until
      %regular monitoring was stopped around day 100. There was a 
       %temporary increase of count rate above 2 cts s$^{-1}$ in the first
        %week of the third month, and the measured count rate started decreasing afterwards.

     \subsection {\xrgs\ spectrum}
\begin{figure}
    \centering
    \includegraphics[width=0.8\textwidth]{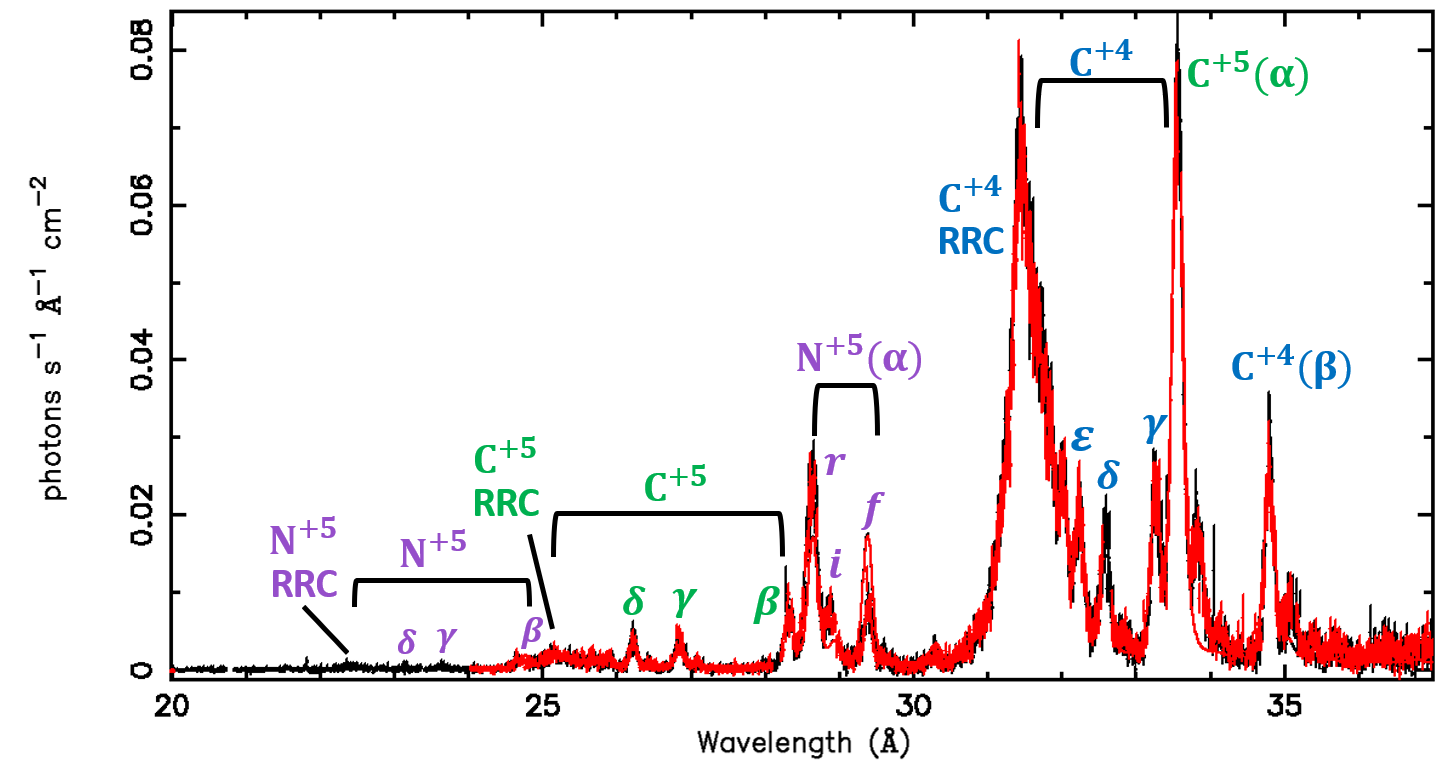}
  \caption{Fluxed spectrum of nova YZ Ret, observed in 2020-09-23 by \xrgs\ 77 days after the eruption. Black and red data are, respectively, \rgs\ 1 and 2. Spectral lines of N$^{+5}$, C$^{+4}$, and C$^{+5}$ are identified. The most prominent spectral feature is the narrow RRC of C$^{+4}$ at 31.5\,\AA , along with the C$^{+4}$ He series %denoted with $\alpha, \beta, \gamma...$% for the transition of n=2,3,4... to n=1 (ground)
  converging to it.  
  The much fainter narrow C$^{+5}$ RRC is just barely observed around 25.2\,\AA , along with the C$^{+5}$ Lyman series. The N$^{+5}$ RRC at 22.3\AA\ and its He series are also detected, and are shown more clearly in panel (a) of Fig.\ref{fig:full}. 
  } \label{fig:yzret}
\end{figure}

     Nova \yzret\ was observed with  \xmm\ 77 days after the initial eruption, between 2020-09-23 13:36 and 2020-09-23 21:22 UT (ObsID 0871010101; PI: K. Sokolovsky) for a total exposure time of 28 ks. 
     The {\sl Swift} XRT count rate during that week was around 2\,cts\,s$^{-1}$, already close to its maximum.
     %The XMM-Newton observatory
We use both \rgs1 and \rgs2 first-order spectra in 
%\citep[RGS,][]{denHerder2001}.
the wavelength range of 5–37\,\AA, corresponding to
the 0.33–2.5\,keV energy range. 
We extracted the spectra
with the XMM-SAS (XMM Science
Analysis System) version 19.1\footnote{\url{https://xmm-tools.cosmos.esa.int/external/xmm_user_support/documentation/sas_usg/USG.pdf}} %We did not use the optical monitor as the target was still too bright, with a visual magnitude $\sim$8.8. The EPIC was operating with the following configuration: pn – small window with thick filter, MOS1 – small window with thick filter, MOS2 – timing with medium filter.
%We extracted the RGS first-order spectra with the XMM-SAS task 
{\it rgsproc} task.
Periods of high background were rejected. 
As explained in detail in \citet{Sokolovsky2022}, we corrected the \rgs\ spectra for pile up.  
     Fig.\,\ref{fig:yzret} shows the spectra observed with \rgs1 and \rgs2. % which are the two identical units equipped on the telescope. 
     
\subsection{\chletg\ spectrum}
\begin{figure}
    \centering
    \includegraphics[width=0.8\textwidth]{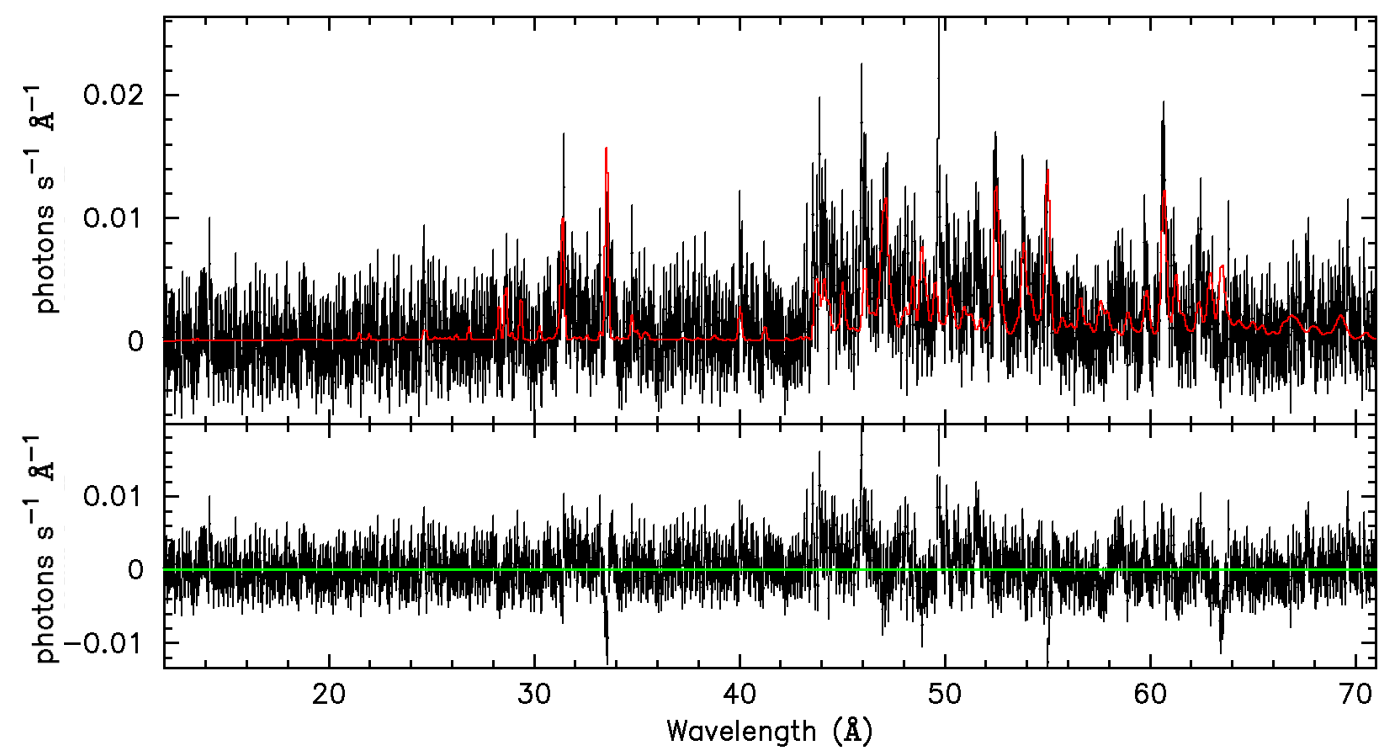}
 \caption{Spectrum of YZ Ret observed on 2020-10-31 for 31 ks with  Chandra and the \letg +HRC-I instrumental setup. 
 For clarity, given the low count rates, the spectrum is presented in photons (not flux), and with the fitted model (Sec.\,\ref{sec:full}) to guide the eye. with residuals plotted in lower panel.} 
 \label{fig:chanyz}
 \end{figure}

\yzret\ was observed with  the \chandra\ \letg\ on 2020-10-31 for 31\,ks.
At the time of the \yzret\ eruption, the High Resolution Camera (HRC)
 aboard {\sl Chandra} was not available in its HRC-S configuration, 
 which is usually used with \letg\ (up to 175\,\AA). 
 Instead,  we used the HRC-I configuration to obtain a spectrum between 1.2 and 
     71 \AA.
%The spectral resolution of the \letg\ is 20 $\times \lambda$ longwards of 3 \AA.
The data were extracted with the
{\sl Chandra} CIAO data analysis package version 4.14.1 and the 
first-order grating redistribution matrix files and ancillary response files, using the CIAO task {\it chandra repro}, with version 4.8.5 of the calibration package CALDB. 
We summed the  positive- and negative- first order spectra with the {\it combine grating spectra} CIAO task, to increase the signal-to-noise ratio (S/N). We did not correct for the higher spectral orders, which have a negligible contribution. 
\yzret\ is by far fainter during this epoch than during the \xmm\ observation.
Fig.\,\ref{fig:chanyz} shows the \letg\ spectrum. 
 %\FloatBarrier

\section{Diagnostic methods for high resolution spectra} 
%\section{Method: plasma diagnostics} 
\label{sec:analysis}

In order to properly model the rich line resolved spectra, we employ a collisional-radiative (CR) model. Assuming the ions are collisionally-ionized, excited by electrons, and in steady-state, the model computes level populations, and predicts the
emission spectral line intensities (in units of ph\,s$^{-1}$cm$^{-3}$) associated with each transition between bound levels $j$ and $i$ ($j>i$) at the source 

 \begin{equation}
    \label{eq:emiss}
   I_{ji}=n_jA_{ji}
\end{equation}

\noindent where $n_j$\,(cm$^{-3}$) is the number density of ions in level $j$ and $A_{ji}$\,(s$^{-1})$ is the Einstein coefficient for spontaneous radiative decay. 
The level populations generally depend on electron density $n_e$ and temperature $T_e$. 
At sufficiently high density, electron impact transitions between excited levels can change the relative populations ($n_j$ in Eq.\,\ref{eq:emiss}), hence the density sensitivity and diagnostics.
%In novae, the plasma may not be  exclusively collisionally ionized. The central supersoft X-ray source may photo-ionize it \citep{***}. The ionization state of the observed plasma will then will not depend directly on $T_e$, but on the ionization parameter $\xi$ \citep[in cgs units of erg\,cm\,s$^{-1}$,][]{Tarter69}
%\begin{equation}
%    \label{eq:xi}
%   \xi=\frac{L}{n_e r^2}
%\end{equation}
%\noindent where $r$\,(cm) is the distance
%from the WD to the observed plasma, and $L$\,(erg s$^{-1}$) is its luminosity. 
%This adds a degree of complexity to the CR model, as photo-excitation (PE) and photo-ionization (PI) processes need to be included, and their rates depend on (the generally unknown) distance $r$, through the WD photon flux $\propto L/r^{-2}$.
%On the other hand, if $\xi$ is estimated from the observed ion and detailed balance computations, while $L$ is measured, the density diagnostic reveals $r$.
%Most generally, in CR models of photoionized plasmas $n_e$ and $r$ diagnostics cannot be disentangled \citep{Peretz2019, Mitrani23}. 
%In order to self consistently include electron- and photon- impact processes, one needs to assume a value for $\xi$ (Eq.\,\ref{eq:xi}) that relates $n_e$ to the photon flux. 
%In the current \rgs\ spectrum of \yzret\ no X-ray ionizing continuum is observed. It is thus unclear a-priori whether the plasma is photoionized or not.
The populations of all levels of a given ion are obtained by solving rate equations that include all of the processes that populate and depopulate each level \citep{Mitrani23}. Here, we add ionization and recombination processes. The atomic structure and coefficients were computed using the relativistic Hebrew University Lawrence Livermore Atomic Code \citep[HULLAC,][]{Bar-Shalom2001}. 

%\textcolor{red}{I think this is enough here.}
%The level $i$ population $n_i$ is solved from a set of linear rate equations:
%%%
%\begin{equation} \label{eq:rate_eq}
%    \begin{split}
%        \frac{dn_i}{dt}=0 &=\sum_{j<i} \left( n_j n_e %Q_{ji}+n_j R_{ji}^{PE} \right) +\sum_{k>i} \left( n_k n_e %Q'_{ki}+n_k A_{ki}\right)\\
%        &-\sum_{j < i} \left( n_i n_e Q'_{ij}+ n_i A_{ij} \right) -\sum_{k > i} \left( n_i n_e Q_{ik}+n_i R_{ik}^{PE} \right)\\
%        &+ n_{+} n_e^2 \alpha^{TBR}_i+n_{+} n_e \alpha^{RR}_i - n_i n_e S_i^{CI} -n_i R_i^{PI}
%    \end{split}
%\end{equation}

%\noindent where $n_{+}$ is the number density of the ion in the higher ionization degree. The first and second terms represent the populating processes of level $i$ from lower and higher levels. The third and fourth terms, represent the depopulating processes of level $i$ to lower and higher levels. The last four terms represent the ionization processes.
%In a steady state, i.e. $dn_i/dt=0$, Eq.\,\ref{eq:rate_eq} reduces to a set of linear algebraic equations, which is solved to get the set of population densities \{$n_i$\}. In order to consistently include both CE and PE, one needs to assume a value for the ionization parameter $\xi$ (Eq.\,\ref{eq:xi}) that relates $n_e$ to the photon flux.

\subsection{Narrow Radiative Recombination Continua (RRCs)} \label{subsec:rrc}
%The \rgs\ spectrum of \yzret\ provides two sensitive electron density diagnostics.
%The more common one is the He-like triplet of N$^{+5}$, to be revisited below.
%A new method, introduced here, utilizes the Rydberg series of lines leading to the narrow radiative recombination (RRC) edge of C$^{+4}$.
\subsubsection{$T_e$ Diagnostics}
The free-bound transition of a plasma electron recombining with an ion emits a photon with an energy that equals the kinetic energy of the electron plus the ionization energy $\rm E_{ion}$ (of the recombined species). The emitted photons
%in the RRC 
delineate the energy distribution of the electrons in the plasma, e.g., a Maxwell-Boltzmann (MB) distribution.
Since the emitted photons have a sharp minimum energy of $\rm E_{ion}$, the radiative recombination continuum (RRC) features a sharp edge at this energy, and usually a MB distribution thereafter:
%The velocities (and thus, the kinetic energy) of free electrons in plasma usually have a Maxwell-Boltzmann distribution which depends on the plasma temperature. Thus, the emitted photons will form a continuous spectral feature over a range of energies with a sharp edge on the ionization energy of the ion. This feature is called the Radiative Recombination Continuum (RRC). 
%The RRC is a highly sensitive to the gas temperature, which can be measured by fitting the RRC model:

\begin{equation}
    f(E)= 
\begin{cases} \label{eq:rrc_mo}
    0,& E<\rm{E}_{ion}\\
    (C/kT_e)e^{-\frac{(E-\rm{E}_{ion})}{kT_e}},  & E \geq \rm{E}_{ion}
\end{cases}
\end{equation}

\noindent where $C$ %\,(photons cm$^{-2}$ s$^{-1}$) 
%Sharon, note the division by kT, so need eV/cm^2/s, but anyhow so far this is very general and does not require units yet
is a flux normalization factor.

In hot plasmas ($T_e>10^5$\,K), the RRC is significantly broadened  by the temperature, %due to thermal broadening. 
and is hard to discriminate from other continuua, primarily bremsstrahlung.
When the plasma is colder ($T_e<10^5$\,K),
narrow RRCs can be discerned in the spectrum, and their width provides exquisite $T_e$ diagnostics.
Narrow RRCs are routinely observed in photoionized plasmas \citep{Liedahl96, Kinkhabwala2002}, where the electrons remain cold despite the high-energy ionizing photons and the high ionization degree of the plasma.
However, narrow RRCs were also observed in a few hot, shocked X-ray sources, where no photo-ionizing continuum was observed nor expected \citep[][]{Schild04, Nordon2009, Yamaguchi2009}.
In nova V4743 Sgr, a bright C$^{+4}$ RRC was observed during the obscuration of the SSS \citep[][Fig.\,4 therein]{Ness2003}, but not identified.

%This was observed, for example, in the Suzaku spectrum of the SNR IC 443 \citep{Yamaguchi2009}. At lower temperatures, the bremsstrahlung contribution is much smaller, the RRC is narrow and can be discerned more clearly.

%Sharon, let's leave this for the discussion
%In transient events such as supernova and nova explosions, the gas is shock-heated to high temperatures and thus ionized. The RRC feature in the spectrum can be amplified due to a rapid expansion and cooling of the plasma. 
\subsubsection{High-$n$ Rydberg Series}
When the RRC is sufficiently bright and narrow, and the resolution of the spectrometer sufficiently high, one can hope to resolve the Rydberg series of high-$n$ (principal quantum number) emission lines converging onto the RRC. We are aware of one such observation to date \citep{Nordon2009}, and the second one is the present \rgs\ spectrum of \yzret .  
If observed, this line series enables an interesting density diagnostic. 
The branching ratio for 
radiative decay from high-$n$ levels, e.g., to the ground level (i.e., the observed X-ray line) is suppressed by collisional ionization, for which the rates increase with $n$.
Thus, at some $n_{\rm max}$ value, ionization dominates over line emission and the series is truncated. 
This effect of continuum lowering is known since the early days of quantum mechanics \citep{Dewey1928, Inglis1939}. Since collisional ionization rates ($S^{CI}$) increase with $n_e$ and radiative decays ($A$) do not, 
the observed $n_{\rm max}$, or equivalently the effective widening of the RRC edge, is a sensitive diagnostic of $n_e$:

\begin{equation} \label{eq:rrc_density}
    \frac{A_{n_{\rm max}+1,1}}{S^{CI}_{n_{\rm max}+1}} < n_e < \frac{A_{n_{\rm max},1}}{S^{CI}_{n_{\rm max}}}
\end{equation}

High-$n$ levels are broadened by the Stark effect, which depends strongly on the plasma (ion) density. 
This makes the high order lines blend with the RRC above some $n_{\rm max}$ that depends on ion density, assumed to be $\sim n_e$. 
\citet{Inglis1939} give the following expression for the dependence of density on $n_{max}$ in the Stark effect:

\begin{equation} \label{eq:stark}
    n_e=0.027n_{max}^{-7.5}a_0^{-3}
\end{equation}
where $a_0=0.529\times10^{-8}$\,cm is the Bohr radius.

\subsection{He-like lines} \label{subsec:helike}

The most common density diagnostic X-ray lines are those of the so-called He-like triplets, whose relative intensities depend on the density and temperature of the plasma \citep{Gabriel1969}. 
%It is widely used as a density diagnostic for CI plasma such as the solar corona and the ISM. 
%These close spectral lines originate from radiative decays (Eq.\,\ref{eq:emiss}) from the three 1s$nl\,(J=1)$ levels ($l=$ s, p) to the ground level 1s$^2\,(J=0)$. Here, $n$ is the principal quantum number of the second electron, and $J$ is the total angular momentum of the level.
The three transitions from the 1s2s,   1s2p$_{3/2}$, and  1s2p$_{1/2}$ $J=1$ levels to the ground level 1s$^2\,(J=0)$ are referred to, respectively, as the  \textit{forbidden} ($f$),  \textit{intercombination} ($i$), and \textit{resonance} ($r$) lines. 
The 1s2s $(J=1)$ level is a meta-stable level with a relatively long life time. Hence, the line ratio 

\begin{equation}
    \label{eq:R_density}
   R(n_e)=\frac{f}{i}
\end{equation}

\noindent is sensitive to $n_e$ in a defined range where collisional de-excitation rates are comparable to the (low) decay rate $A_{ji}$ from that level. These density ranges in He-like ions increase with the atomic number, as $A_{ji}$ increases with the nuclear charge \citep{Gabriel1969}.
Using the CR model we 
%use Eq.\,\ref{eq:rate_eq} to 
compute the expected ratios as a function of $n_e$, and infer the density by comparing with the measured ratio.

\subsection{Emission Measure} \label{subsec:em}

The emitted flux from an optically thin uniform plasma is proportional to the emission measure $EM=\int n_e n_H dV \approx n_e n_H V$ where $n_H$ (cm$^{-3}$) is the hydrogen density, and $V$ (cm$^{3}$) is the volume of the emitting plasma. An $EM$ can be obtained directly from discrete features, in this case the RRC photon luminosity $L_{RR}=n_q n_e \alpha^{RR} V$, where $\alpha^{RR}$ is the RR rate coefficient, and $n_q$ is the number density of the recombining ion $q$. Since $n_q=f_q A_Z n_H$, the photon flux can be written as:
\begin{equation} \label{eq:emission_meas_RRC}
    F^{RRC} = \frac{1}{4\pi d^2} f_q A_Z \alpha^{RR} EM
\end{equation}
where $f_q$ is the ion fraction, $A_Z$ is the elemental abundance, and $d$ is the distance to the source.
%In a collisional plasma, when observing a spectral line of transition $j$ to $i$,  V$, 
%In a steady state, $n_j A_{ji}=n_e n_j S_{ij}$ because the excitation and decay are balanced, %(usually referred to as detailed balance), 
%and thus the emission line flux 
%$F_{ji}\:(\rm{ph\,s^{-1}cm}^{-2})$ is (assuming a uniform density):
%\begin{equation} %\label{eq:emission_meas_flux}
%    F_{ji}^{line} = \frac{L_{ji}}{4\pi d^2}
%    = \frac{1}{4\pi d^2}  \frac{n_i}{n_q} f_q A_Z S_{ij} n_e n_H V \propto n_e n_H V = EM
%\end{equation}
%where $n_q,n_i$ (cm$^{-3}$) are the number density of the ion, and the ion in excited state $i$, respectively. $f_q$ is the ion fraction with respect to all other ions of the same element. $A_Z=n_Z/n_H$ is the abundance with respect to hydrogen where $n_H$ is the hydrogen number density. 
%In fully ionized plasma, one can further assume $n_e\cong 1.2n_H$. Thus, the emission measure is  approximately:
%\begin{equation} %\label{eq:emission_meas}
%    EM= n_e^2 V
%\end{equation}
In a hot collisionally-ionized plasma, where all photons are emitted following electron-ion collisions, the photon flux $F$ is exactly proportional to the $EM$. 
Thus, by measuring the flux, one obtains the $EM$.
In the Xspec \citep{Arnaud1996} model we use in Sec.\,\ref{sec:full} \citep[\textit{bvapec,}][]{Smith2001}, $F$ is parameterized  by:
%(\citep{Foster2012,Leahy2023}):

\begin{equation} \label{eq:emission_meas_vapec}
    F = \frac{10^{-14}}{4\pi d^2}EM
\end{equation}

\noindent Once $d$ is known, and the density is measured, the $EM$ reveals the emitting volume $V$. 

%\clearpage

\section{Analysis of XMM-Newton RGS spectrum} 
%\section{Results} 
\label{sec:spec}

The \rgs\ spectrum allows us to derive the temperatures, densities, and EM of the X-ray source at day 77. The Lyman, and He series of C$^{+5}$ and C$^{+4}$, as well as that of N$^{+5}$, are clearly identified, all blueshifted from their laboratory wavelengths. The predominant feature in the spectrum is the narrow RRC of C$^{+4}$ around 31.5\,\AA , which is broadened and asymmetric due to the electron temperature on the short-wavelength side, and the high-$n$ lines on its long-wavelength side. 
Similar but much fainter features are the narrow C$^{+5}$ and N$^{+5}$ RRCs observed around 25.2\AA\ and 22.3\AA, respectively. The He-like triplet of N$^{+5}$ is observed around 29\AA , and the He-like triplet of C$^{+4}$ is at 40\AA\ outside the \rgs\ band. 

%Sharon, this goes in the discussion, but first what was measured.
%The high ionization degrees observed in the spectrum can be obtained by collisional processes if the temperature of the object is $kT\sim100-1000$\,eV, which is higher than the measured temperature of $kT=2\,$eV. Radiation processes can produce cold plasmas with high ionization degrees, but the spectrum shows no continuum.

\subsection{C$^{+4}$, C$^{+5}$ and N$^{+5}$ RRCs} 
\label{subsec:RRCdiagnostics}
\begin{figure}[h]
    \centering
    \includegraphics[width=0.6\textwidth]{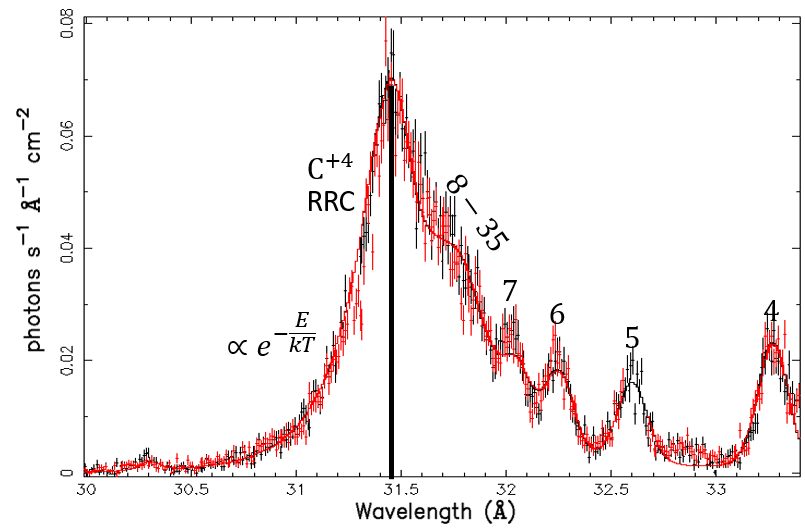}
  \caption{The 2020 \rgs\ spectrum of \yzret\ in the range 30-33.5\,\AA\ highlighting the C$^{+4}$ RRC region. %, along with a series of high-$n$ C$^{+4}$ converging to it.  
  Black and red data are \rgs\ 1, and 2, respectively.
  The solid line is a model fitting the RRC, and the $n$ to 1 Rydberg series up to $n_{\rm max}$ as gaussians. The RRC exponential width on its blue side indicates $kT_{e,C^{+4}}=1.96\pm0.04$\,eV. %Line series were constrained to have amplitudes that decrease as $n^{-3}$. 
  The best-fit $n_{\rm max}$ is 35 (see Fig.\,\ref{fig:cstatmax}). 
} \label{fig:rrc35}
\end{figure}

\begin{figure}[h]
    \centering
    \includegraphics[width=0.47\textwidth]{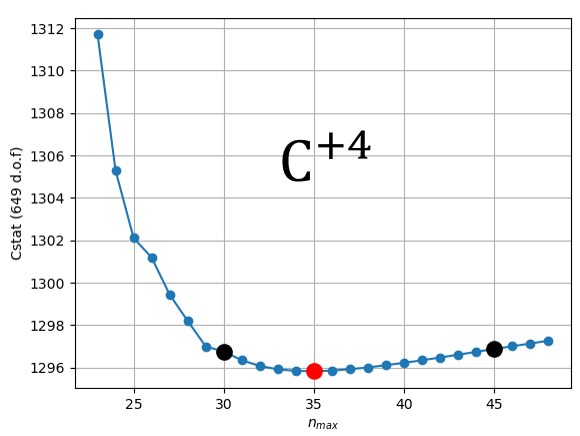}
    \includegraphics[width=0.47\textwidth]{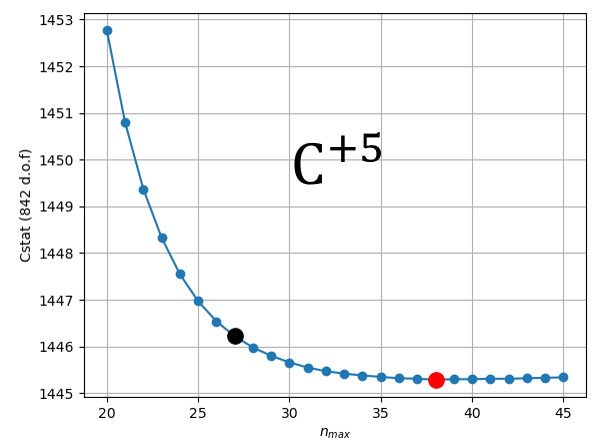}
  \caption{
  Values of $C_{\rm stat}$ obtained from fitting the Rydberg series up to $n_{\rm max}$, the maximal $n\rightarrow1$ transition included in the model for C$^{+4}$ (left) and C$^{+5}$ (right). 
  The minimal $C_{\rm stat}$ value (red dot) and uncertainty region of $\Delta C_{\rm stat} = 1$ (black dots) yield $n_{\rm max}=35^{+10}_{-5}$ for C$^{+4}$, and $n_{\rm max}=38(>27)$ for C$^{+5}$, which from Eq.\,\ref{eq:rrc_density} indicates an electron density of $n_{e,RRC}=(2.4^{+3.0}_{-1.8})\times10^{11}$\,cm$^{-3}$ (for C$^{+4}$) and $n_{e,RRC}=(1.6(<9.6))\times10^{11}$\,cm$^{-3}$ (for C$^{+5}$) at 90\% confidence.} \label{fig:cstatmax}
\end{figure}

A zoom into the C$^{+4}$ RRC region between 30-33.5\,\AA\ is shown in Fig.\,\ref{fig:rrc35}, with the bright RRC blueshifted by $\sim1500$\,km\,s$^{-1}$ and peaking at 31.45\,\AA .
%The short-wavelength exponential profile of the RRC is due to the (Maxwellian) electron temperature.
%The appreciably broad shoulder on the long-wavelength side can be explained by a Rydberg series of emission lines that are emitted by decays from high-$n$ levels to the ground $n=1$ level. 
We modeled this region with an RRC (Eq.\,\ref{eq:rrc_mo}) and a series of gaussians representing the Rydberg series. 
The RRC model provides a temperature measurement of $kT_{e,C^{+4}}$=$(1.96\pm0.04)$\,eV. The series of unresolved C$^{+4}$ lines are fixed to the theoretical wavelengths of the $np$ to $1s$ transitions, and their Gaussian width to that of the global fit $\sigma_v=520$\,km\,s$^{-1}$ (Sec.\,\ref{sec:full}). Line intensities for $n\geq10$ are scaled as $n^{-3}$, to follow the hydrogenic scaling of the Einstein coefficients. 
This is a good approximation for high-$n$ levels, which are populated according to the Boltzmann distribution, and their close proximity implies comparable populations. 
The best fit to the spectrum is obtained when the highest level transition is $n_{\rm max}=35$. In Fig.\,\ref{fig:cstatmax} (left panel), we plot the Cstat ($C$) fitting parameter as a function of $n_{\rm max}$, which demonstrates that $\Delta C_{\rm stat} = 1$ corresponds to $n_{\rm max}=35^{+10}_{-5}$. 
Performing the same fit with $\chi^2$ yields $n_{max}=34^{+8}_{-5}$ with 90\% confidence.

In order to measure $n_e$ from $n_{\rm max}$, we need to scale $A$ and $S^{CI}$ with $n$ (see Eq.\,\ref{eq:rrc_density}).  
We extrapolated the HULLAC values for $n\geq10$ according to $A_{n\rightarrow 1}\propto n^{-3}$, $S^{CI}_{n\rightarrow \infty}\propto n^2\log(n)$ \citep{Lotz1968}, which implies (c.f., Eq.\,\ref{eq:rrc_density}):

\begin{equation}
    \label{eq:scaling}
   n_e\simeq \frac{A_{n_{\rm max}\rightarrow 1}}{S^{CI}_{n_{\rm max}\rightarrow \infty}} \propto \frac{n_{\rm max}^{-5}}{\log(n_{\rm max})}
\end{equation}

\noindent The inferred $n_e$ from $n_{\rm max}=35^{+10}_{-5}$ is $n_{e,RRC}=(2.4^{+3.0}_{-1.8})\times10^{11}$\,cm$^{-3}$. Such high densities occur either in the dense blobs of ejecta, or in the accretion disk. Similarly high density has been found in Chandra X-ray grating spectroscopy of nova V959 Mon \citep{Peretz2016}.

The same diagnostics were applied to the C$^{+5}$ RRC, which is observed at 25.2\AA. The short-wavelength exponential profile yields a temperature measurement of $kT_{e,C^{+5}}=1.9\pm0.2$\,eV, consistent with the C$^{+4}$ temperature. The long-wavelength side of the C$^{+5}$ RRC was fitted with a high-$n$ levels series of Gaussians as the C$^{+4}$ RRC yielding $n_{\rm max}=38$. Fig.\,\ref{fig:cstatmax} (right panel) shows that $n_{\rm max}$ can only be constrained from below  $n_{\rm max}>27$. Using Eq.\,\ref{eq:scaling} we infer a density of $n_{e,RRC}<9.6\times10^{11}$\,cm$^{-3}$ (upper limit), with a best fit of  $n_{e,RRC}=1.6\times10^{11}$\,cm$^{-3}$.

The same procedure was applied to the N$^{+5}$ RRC, which only provides an upper limit of $kT_{e,N^{+5}}<1.5$\,eV. The low flux of the N$^{+5}$ RRC does not allow fitting of the high-$n$ series and thus, no density diagnostic.

Interestingly, if we plug in the best constrained $n_{\rm max}=35$ from C$^{+4}$ into the Stark-broadening density diagnostic (Eq.\,\ref{eq:stark}), we 
obtain $n_{e,Stark}=4.8\times10^{11}$\,cm$^{-3}$, which is 
within the uncertainties of the collisional continuum lowering estimate of Eq.\,\ref{eq:scaling}.
In summary, the RRCs of \yzret\ at day 77 indicate the presence of $kT_e \approx 2$\,eV electrons with an electron density in the range of $1-5\times 10^{11}$\,cm$^{-3}$.

%compute the electron density $n_{e,Stark}=(2.6(<33.6))\times10^{11}$\,cm$^{-3}$, which is of the same order of magnitude. By using Eq.\,\ref{eq:stark} that infer the electron density due to the Stark effect, we compute the electron density $n_{e,Stark}=(5^{+10}_{-4})\times10^{11}$\,cm$^{-3}$. This value is consistent with the collisional effect computation.

\subsection{Charge Exchange}
\label{sec:cx}
How exactly the high-$n$ levels get so highly populated in \yzret\ and their lines so intense is an intriguing question.
In fact, to the best of our knowledge, such a spectrum of C$^{+4}$ has never been observed, including not in the laboratory.
The high intensity of high-$n$ lines is a smoking gun of charge exchange (CX). 
However, due to energy conservation, CX is highly selective in its population of excited levels. 
%The maximal population of 
C$^{+5}$ exchanging charge with a neutral H atom, both in the ground state, results primarily in C$^{+4}$ ions in the $n=4$ levels, with $n=3, 5$ levels being appreciably less populated \citep{Nolte2012}.
High collision velocities could result in somewhat higher levels, but equal fluxes of $n = 4 -7$ lines  and beyond as observed in Fig.\,\ref{fig:rrc35} can not be explained by a high velocity alone.
Conversely, CX with {\it excited} H atoms (or of excited C$^{+5}$ in the initial state) requires much less energy to pull out the bound electron, and thus can end in higher-$n$ levels of C$^{+4}$ \citep[e.g.,][]{Janev1985}.
Since excited levels are close to each other, CX with H in (multiple) excited levels will populate a broader distribution of C$^{+4}$ high-$n$ levels, as observed here.

We therefore suggest CX in \yzret\ is occurring with excited H. At the measured electron density in the ionized gas of $n_e \approx 10^{11}$\,cm$^{-3}$, $n=2$ levels would be populated to 0.2\% of the ground level.
We have no measure of the neutral gas density, which could be denser than the ionized one.
We conclude that the high-$n$ lines of C$^{+4}$ in the \rgs\ spectrum of \yzret\ feature the first unambiguous, and most conspicuous evidence for CX in an astrophysical plasma (at least aside from the solar wind).  
Moreover, CX is completely in line with the observation of highly charged ions mixing with cold gas, as evident also by the RRCs.
%As the velocity of the CX collision increases, so does the $n$ level to which capture occurs. Population of $n=8$ for C+4 require a few 1000 km/s, after which the cross section for all $n$-levels dcreases sharply. 
%https://dbshino.nifs.ac.jp/nifsdb/chart/top

\subsection{He-like N$^{+5}$ lines} 
\label{subsec:Ndiagnostics}

\begin{figure}[h]
    \centering
    \includegraphics[width=0.43\textwidth]{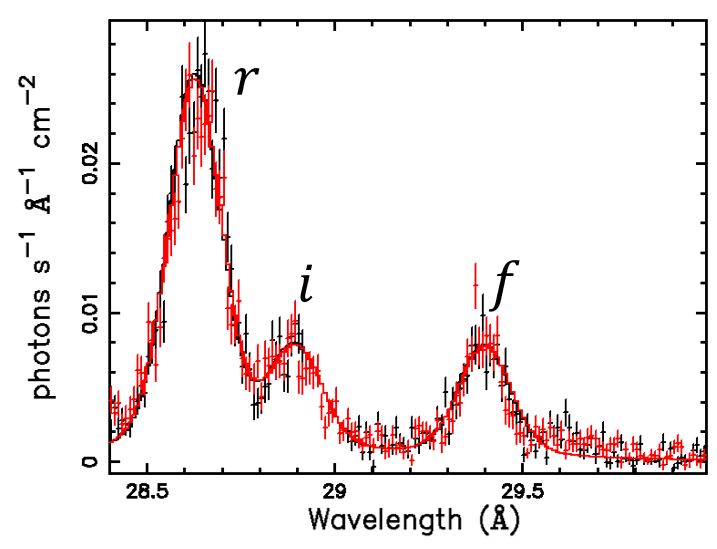}

  \caption{The $r, i, f$ lines of N$^{+5}$ observed in the YZ Ret \rgs\ spectrum. The measured $f/i$ ratio is $1.1\pm0.2$. 
  }
  \label{fig:n+5}
\end{figure}

In this section, we perform a direct measurement of the N$^{+5}$ He-like line intensities, in order to infer the density of the plasma by using Eq.\ref{eq:R_density}. Fig.\ref{fig:n+5} shows the three observed spectral lines of He-like N$^{+5}$, fitted with three Gaussians. The measured wavelength and flux of each line is presented in Table.\,\ref{tb:n5_2}, and the measured $f/i$ ratio is $R(n_e)_{\rm{N}^{+5}}=1.1\pm0.2$ (Eq.\,\ref{eq:R_density}). 
The average blueshift of the three lines is $v=-1620\pm80$\,km/s, which is slightly higher than the global fit value in Sec.\,\ref{sec:full}. 
The discrepancy can be due to an asymmetry of the $r$ line, and 1s$2l2l'$ satellite lines that blend with the N$^{+5}$ lines, mostly the $i$ line, similar to the $2l2l'$ satellites of C$^{+5}$ at 33.85\AA\ (Sec.\ref{sec:dr}). 

\begin{figure}
    \centering
    \includegraphics[width=0.7\textwidth]{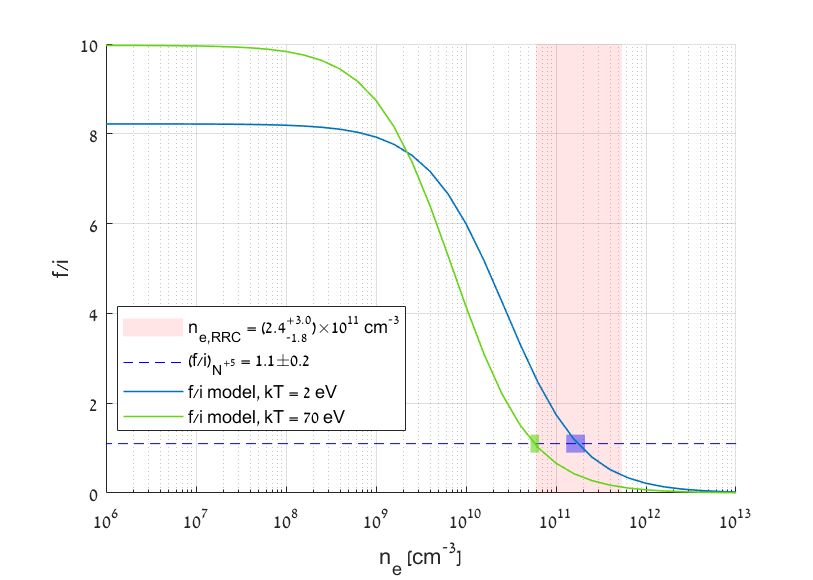}
  \caption{Computed $f/i$ ratio as a function of $n_e$ for $kT_e=2$\,eV (blue) and 70\,eV (green).  The blue dashed line is the measured $f/i$ ratio in the spectrum. Shaded areas indicate uncertainties. The measured ratio indicates a density of $n_e=(1.7\pm0.4)\times10^{11}$\,cm$^{-3}$, for 2\,eV, fully consistent with $n_{e,RRC}=(2.4^{+3.0}_{-1.8})\times10^{11}$\,cm$^{-3}$ (Sec.\ref{subsec:RRCdiagnostics}), which is represented by the red shaded area.
  } \label{fig:n+5model}
\end{figure}

We solve the rate equations (Sec.\,\ref{sec:analysis}) to obtain the theoretical ratio at $kT_e$=2\,eV (the cold recombining plasma component described in Sec.\,\ref{subsec:RRCdiagnostics}) and at 70\,eV (the hot plasma component described in Sec.\,\ref{sec:full}) to cover the expected temperature range. The computed $f/i$ ratio as a function of $n_e$ is presented in Fig.\,\ref{fig:n+5model}. The measured value of $R(n_e)_{\rm{N}^{+5}}=1.1\pm0.2$ indicates $n_e=(1.7\pm0.4)\times10^{11}$\,cm$^{-3}$, and $n_e=(5.8\pm0.7)\times10^{10}$\,cm$^{-3}$ for 2\,eV and 70\,eV, respectively. If satellite lines indeed blend with the $i$ line, $R(n_e)$ is likely higher and the density is slightly lower (Fig.\,\ref{fig:n+5model}). The density for 2\,eV is consistent with the value we obtained from $n_{\rm max}$ at the RRC $n_{e,RRC}=(2.4^{+3.0}_{-1.8})\times10^{11}$\,cm$^{-3}$, and a factor 3 higher than the 70\,eV value. The above measurement does not include photo-excitation (PE) because there is no observed continuum in the \yzret\ spectra.%, and the gas does not appear to be photo-ionized.

We check whether the gas may be photo-ionized and PE processes could dramatically change our density measurement. % by adding PE to the rate equations. 
The unobserved ionizing source we hypothesize is a continuum black body with a temperature of $kT_{bb}$=50\,eV and a luminosity of $L=10^{38}$\,erg\,s$^{-1}$, high but typical for a SSS. In photo-ionization steady state, the PE flux and $n_e$ are related through the ionization parameter $\xi=L/(n_e r^2)$, where $r$ is the distance from the ionizing source to the emitting plasma. N$^{+5}$ typically forms at $\xi=10$\,erg\,s$^{-1}$cm \citep{Kallman2009}.
Adding PE processes slightly reduces the density by 20\%; at $kT_e=2$\,eV, $n_e=(1.4\pm0.4)\times10^{11}$\,cm$^{-3}$. $\xi=10$\,erg\,s$^{-1}$cm then implies $r=(7.6\pm0.9)\times10^{12}$\,cm.
The outflow moving at 1500\,km\,s$^{-1}$ will reach a much larger distance at its observed age of 77 days. In addition, the second observation by \chandra\ shows the outflow at a similar ionization degree, which would be different if it moved away and was photo-ionized by the same SSS flux (see discussion in Sec.\,\ref{sec:discussion}).

%This distance is too low for the 1500\,km\,s$^{-1}$ outflow of \yzret\ at its observed age, let alone if it continues to move out (see discussion in Sec.\,\ref{sec:discussion}).% which rules out a photo-ionized plasma.

%\begin{deluxetable}{ccccccc}[h]
%\tablewidth{0pt}
%\label{tb:n5}
%\tablecaption{Measurements of the three He-like N$^{+5}$ $n=2\rightarrow1$ transition lines. The average blueshift velocity is $v=-1619\pm84$\,km/s.}
%\tablehead{Line&Transition&$\lambda_{meas.}$&$\lambda_{theory}^{(a)}$&$\Delta\lambda$&Velocity\,[km/s]&Flux\,[10$^{-3}$ cts\,s$^{-1}$ cm$^{-2}$]
%}
%\startdata
%{r} & $1s2p_{1/2}\rightarrow1s^{2}$ & $28.63\pm0.01$ &$28.787\pm0.001$&$-0.15\pm0.01$&$-1635\pm104$ &$(5.5\pm0.2)$\\
%{i} & $1s2p_{3/2}\rightarrow1s^{2}$ & $28.91\pm0.02$ &$29.081\pm0.001$&$-0.17\pm0.02$&$-1762\pm206$ &$(1.4\pm0.2)$\\
%{f} & $1s2s\rightarrow1s^{2}$ & $29.39\pm0.01$ &$29.534\pm0.001$&$-0.14\pm0.01$&$-1461\pm102$ &$(1.6\pm0.2)$\\
%\enddata
%\tablecomments{\\ (a) \citet{Cann1992,Drake1969,Drake1971}, taken from the NIST atomic database  \\
%}
%\end{deluxetable}

\begin{deluxetable}{ccccccc}[h]
\tabletypesize{\footnotesize}
\tablewidth{0pt}
\label{tb:n5_2}
\tablecaption{Measurements of the three He-like N$^{+5}$ $n=2\rightarrow1,J=1\rightarrow0$ transitions. The average blueshift velocity is $v=-1620\pm80$\,km/s.}
\tablehead{Line&Transition&$\lambda_{meas.}$[\AA ]&$\lambda_{rest}^{(a)}$[\AA ]&$\Delta\lambda$[\AA ]&Velocity\,[km/s]&Flux\,[10$^{-3}$ ph\,s$^{-1}$ cm$^{-2}$]
}
\startdata
{$r$} & $1s2p_{1/2}\rightarrow1s^{2}$ & $28.63\pm0.01$ &$28.787\pm0.001$&$-0.15\pm0.01$&$-1640\pm100$ &$(5.5\pm0.2)$\\
{$i$} & $1s2p_{3/2}\rightarrow1s^{2}$ & $28.91\pm0.02$ &$29.081\pm0.001$&$-0.17\pm0.02$&$-1760\pm210$ &$(1.4\pm0.2)$\\
{$f$} & $1s2s\rightarrow1s^{2}$ & $29.39\pm0.01$ &$29.534\pm0.001$&$-0.14\pm0.01$&$-1460\pm100$ &$(1.6\pm0.2)$\\
\enddata
\tablecomments{\\(a) 
\citet{Cann1992,Drake1969,Drake1971}, taken from the NIST atomic database  \\
%NIST atomic database \citep{Cann1992,Drake1969,Drake1971} \\
}
\end{deluxetable}

\subsection{Dielectronic recombination satellites} \label{sec:dr}
\begin{figure}[h]
    \centering
    \includegraphics[width=0.9\textwidth]{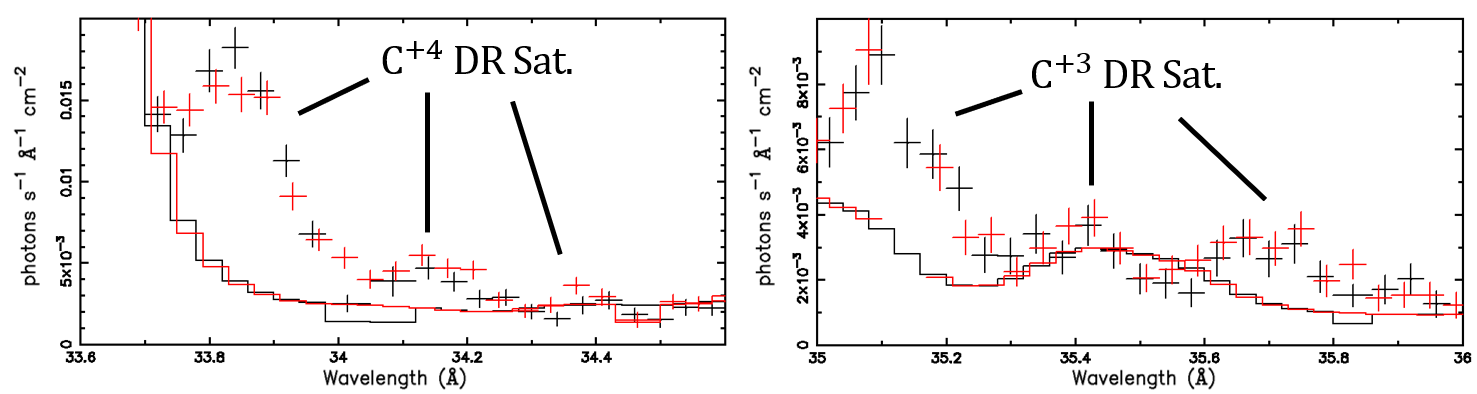}
  \caption{Dielectronic recombination satellite lines of C$^{+4}$ and C$^{+3}$ observed in the \rgs\ 1 and 2 (black, red) spectrum, much above the plasma model prediction (solid line). The data were rebinned for clarity. The C$^{+3}$ satellite lines (right panel) may be blended with Ca$^{+10}$ or Ar$^{+8}$ lines.}
  \label{fig:drsat}
\end{figure}

\begin{figure}
    \centering
    
    \includegraphics[width=0.45\textwidth]{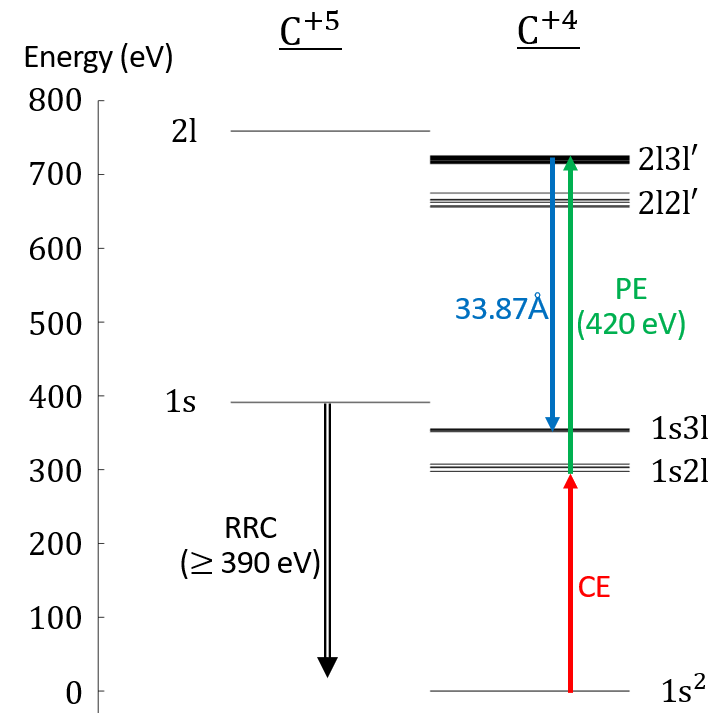}
    \includegraphics[width=0.54\textwidth]{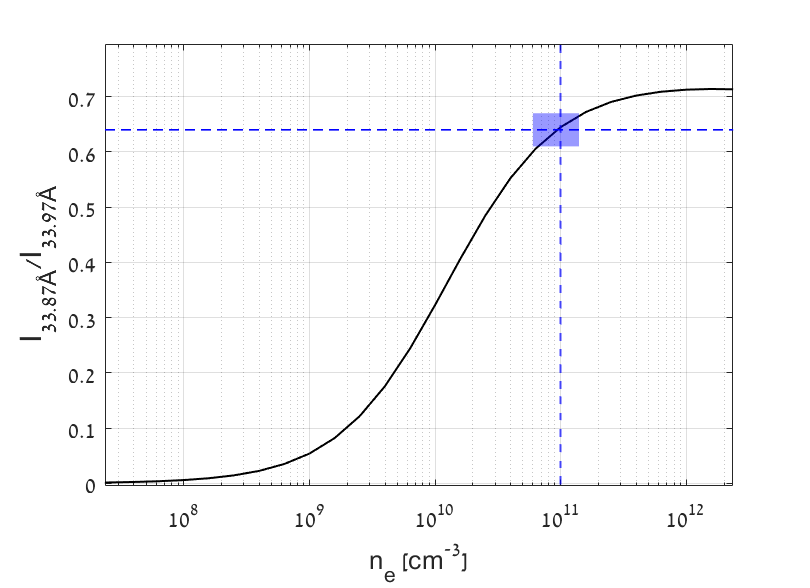}
    
  \caption{\textbf{Left: }Energy diagram of the C$^{+5}$ \& C$^{+4}$ levels that are included in the CR model. 
 %At a density of $n_e=10^{11}$, 
 Ions are excited by CE (red arrow), and further by PE (green arrow) selectively to the $2l3l'$ configuration complex, by C$^{+4}$ RRC photons of 392-420\,eV (29.5-31.6\,\AA ).
  The ions decay to 1s$3l$ levels, which produce the 33.87\AA\ satellite line (blue arrow) that is shown in Fig.\,\ref{fig:synthDR} and in the observed \yzret\ spectrum (Fig.\,\ref{fig:drsat}). 
  \textbf{Right: }Intensity ratio of the 33.87\AA\ satellite line and C$^{+4}$ He$\beta$ at 33.97\AA , as a function of $n_e$.
  Since the population of the 1s$2l$ levels depends on $n_e$, the line ratio provides a density diagnostic of $n_e=(1.0\pm0.4)\times10^{11}$\,cm$^{-3}$, consistent with the other density measurements (Sections \ref{subsec:RRCdiagnostics}, \ref{subsec:Ndiagnostics}).} 
  \label{fig:2l3l}
\end{figure}

Prominent dielectronic recombination (DR) satellite lines of C$^{+4}$ between 33.8-34.4\,\AA , and of C$^{+3}$ between 35-35.8\,\AA\ are observed in the \rgs\ spectrum of \yzret\ (Fig.\ref{fig:drsat}, panel (e) of Fig.\,\ref{fig:full}). These lines exist in the hot collisional component (\textit{bvapec}), but are by far weaker than the observed intensities.
%A significant discrepancy between the \textit{bvapec} model and the observed spectra is due to the %prominent dielectronic recombination (DR) satellite lines of C$^{+4}$ and C$^{+3}$ between 33.8-34.4\,\AA\ and between 35-35.8\,\AA\ (panel (e) of Fig.\ref{fig:full}), respectively. 
The same DR satellites are seen in laboratory plasma, such as laser-produced plasma \citep{Seely1981}, where the electron density exceeds $n_e\simeq10^{20}$\,cm$^{-3}$. %, and the collisions produce these lines. 
In that case, one observes 1s-2p transitions produced by the 2p$2l'$ and 2p$3l'$ doubly excited levels decaying to 1s$n'l'$, at wavelengths above and below 34\,\AA, respectively \citep[c.f. Fig.3 in][]{Seely1981}. 
In contrast, % to the laser produced plasma spectrum, 
in the present \yzret\ spectrum, $n_e$ is much lower, %$%n_e\simeq10^{11}$\,cm$^{-3}$, 
and the $2l3l'$ lines are much brighter than the dim $2l2l'$ lines (Fig.\,\ref{fig:drsat}). These DR satellites are all about 300\,eV above the 1s continuum (see energy diagram in Fig.\,\ref{fig:2l3l}, left panel). 
Therefore, as the temperature approaches $kT\simeq100$\,eV, which is sufficient for dielectronic capture, they should all be populated, and appear in the spectrum.
The weakness of the 2l$2l'$ DR resonances above 34\,\AA\ in the \yzret\ spectrum compared to the bright 2l$3l'$ ones just below 34\,\AA\ beg for an explanation.

The peculiar satellites intensity ratios can be explained by a different mechanism than DR, which is {\it selective} PE photo-pumping from the 1s$2l$ levels to the $2l3l'$ levels, by the intense C$^{+4}$ RRC flux in the narrow band of 392-420\,eV (29.5-31.6\,\AA , Fig.\,\ref{fig:full}, panel (d)). This selective photo-pumping to $2l3l'$ explains why the lower $2l2l'$ levels are not populated (Fig.\,\ref{fig:2l3l}, left panel). 
For this mechanism to operate, the 1s$2l$ levels need to be sufficiently populated, namely $n_e$ needs to be sufficiently high.
At a density of $n_e\simeq10^{11}$\,cm$^{-3}$, the 1s2s (J=1) excited level is populated to 0.2\% of the ground level by CE processes, to be subsequently photo-excited to the $2l3l'$ doubly excited levels. The density dependence of the population of this level provides another diagnostic, pointing yet again to $n_e\simeq10^{11}$\,cm$^{-3}$, see right panel of Fig.\,\ref{fig:2l3l}.
The diagnostic was done with the line ratio of 33.87\AA(satellite)/33.97(C$^{+4}\beta$), but it's not exclusive to the C$^{+4} \beta$ line. The ratio of the satellite line to C$^{+5}$ Ly$\alpha$ can also be used as a density diagnostic, because the density affects the population of the 1s2l (for C$^{+4}$) and 2l (for C$^{+5}$) in a similar way. The correct line ratios are obtained in our model for a density of $n_e=(1.0\pm0.4)\times10^{11}$\,cm$^{-3}$, in agreement with the density measurements in Sections \ref{subsec:RRCdiagnostics}, \ref{subsec:Ndiagnostics}.

\begin{figure}
    \centering
    \includegraphics[width=\textwidth]{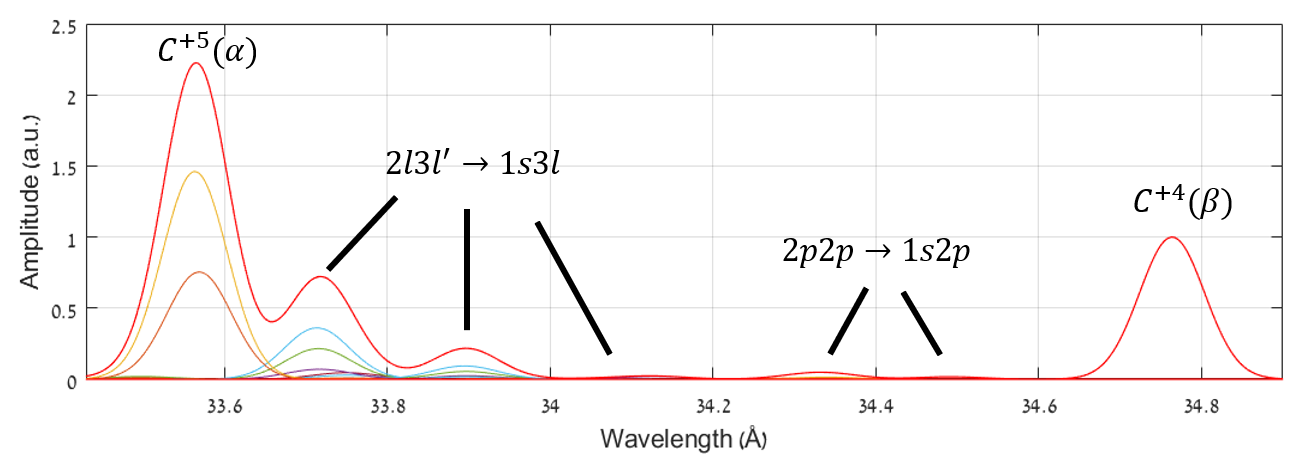}
  \caption{Synthetic spectrum produced by C$^{+5}$ \& C$^{+4}$ CR model with $kT=70$\,eV, and $n_e=10^{11}$\,cm$^{-3}$, including DR via the doubly excited $2l2l'$ and $2l3l'$ configuration complexes, with an addition of PE. The colored lines are the individual transitions while the red line is their sum, which shows line ratios that match the observed spectrum of \yzret. The spectrum is blueshifted by $v=-1500$\,km s$^{-1}$} 
  \label{fig:synthDR}
\end{figure}

The resulting synthetic spectrum of the $kT=70$\,eV CR model is shown in Fig.\,\ref{fig:synthDR}. 
The model includes the DR of C$^{+5}$ via the doubly excited $2l2l'$ and $2l3l'$ configuration complexes, with an addition of photo-exciting flux, with the spectrum of the cold C$^{+4}$ RRC between 392-420\,eV. %We also computed the  DR satellites of C$^{+4}$ via the 1s$2l2l'$ and 1s$2l3l'$ doubly excited configuration complexes. 
The spectrum in Fig.\,\ref{fig:synthDR} was computed at $n_e=10^{11}$\,cm$^{-3}$, and matches the intensities of the satellite lines observed in \yzret .
%The right panel of Fig.\,\ref{fig:2l3l} shows the density dependence of the line ratio between the strongest DR satellite and C$^{+4}$ He$\beta$, at 33.87\AA\ and 33.97\AA\ (rest frame), respectively. 

The presence of a strong satellite line at 33.87\,\AA , along with the absence of strong matching satellite lines at 33.5\,\AA\ and 33.67\,\AA\ (rest frame) thus shows an interaction between the hot (70\,eV) and cold (2\,eV) components of the plasma. The RRC photon flux that is required for PE by the CR model in order to produce the correct intensities of the satellite lines is $F_{\rm RRC}=10^{28}$\,ph\,s$^{-1}$\,cm$^{-2}$. From the distance to YZ Ret of $d=7.8\times10^{21}$\,cm away \citep{Bailer-Jones2021}, we know the RRC luminosity $L_{\rm RRC}$.
If this was a point source, the distance $r$ from it at which $F_{\rm RRC}=L_{\rm RRC}/(4\pi r^2$) is $r=10^7$\,cm.
 This small distance is consistent with the physical picture of a thin shock-heated layer that is discussed in Sec.\,\ref{sec:epoch}, where the ions mix with the cold electrons.

\subsection{Volume and mass of the emitting plasma} 
\label{subsec:volume}
In this section, we use the best-fitted fluxes of the (mixed) cold (RRCs) and hot (Sec.\,\ref{sec:full}) plasmas listed in Table\,\ref{tb:rrc}, to infer their volume from the density measurement (Eqs.\,\ref{eq:emission_meas_RRC}, \ref{eq:emission_meas_vapec}). In all the calculations, we use a distance to \yzret\ of $d=(7.8\pm0.7)\times10^{21}$\,cm \citep{Bailer-Jones2021}, a solar carbon abundance of $A_C=4\times10^{-4}$ \citep{Wilms2000}, and the density value we obtained in Sec.\,\ref{subsec:Ndiagnostics}, $n_e=(1.7\pm0.4)\times10^{11}$\,cm$^{-3}$.

For the cold plasma, we use the radiative recombination rate coefficients at 2\,eV, $\alpha^{RR}_{C^{+5}}=2.33\times10^{-12}$\,cm$^{-3}$\,s$^{-1}$ and $\alpha^{RR}_{C^{+6}}=3.15\times10^{-12}$\,cm$^{-3}$\,s$^{-1}$ (computed by HULLAC) in Eq.\,\ref{eq:emission_meas_RRC} to obtain $f_{C^{+5}}V=(8\pm4)\times10^{35}$\,cm$^{3}$ and $f_{C^{+6}}V=(1.2\pm0.6)\times10^{34}$\,cm$^{3}$. If we assume the C$^{+4}$ and C$^{+5}$ RRCs are emitted from the same volume, then we can infer the ion fraction ratio $f_{C^{+6}}/f_{C^{+5}}=(1.5\pm0.9)\%$, meaning the carbon in the plasma is dominated by the C$^{+5}$ ion ($f_{C^{+5}}\simeq1$), which is consistent with the ionization balance of a hot plasma at a temperature of $kT_e\simeq70$\,eV. Finally, we obtain a volume of $V_{\rm cold}=(8\pm4)\times10^{35}$\,cm$^{3}$.

The volume of the hot plasma (\textit{bvapec} model) is obtained from Eq.\,\ref{eq:emission_meas_vapec}, by assuming the standard $n_e/n_H=1.2$ density ratio for a fully ionized, solar abundance plasma.
%This ratio assumed a nominal hydrogen solar abundance. 
If the hydrogen abundance is different than solar, one needs to introduce a correction factor \citep{Leahy2023}. 
We obtain $V_{\rm{hot}}=(6\pm3)\times10^{35}$\,cm$^{3}$, which is consistent with $V_{\rm cold}$, again lending to the notion that the hot and cold plasmas are mixed.
Given the volume and density, we can estimate the mass in the X-ray source as $M=\mu m_pn_eV $, where $\mu m_p = 1.3m_p$ is the mean atomic mass in the plasma, $m_p$ being the proton mass.
This estimate yields $M=10^{-10}M_\odot$, which is fitting for such a low luminosity source. % with a short cooling time (see Sec.\,\ref{sec:discussion}). 
In Sec.\,\ref{sec:discussion} we show that the mass outflow rate is normal, but that the short cooling times leave only a small mass of hot gas at any given time.

\section{Global fitting}
\label{sec:full}
\subsection{\xrgs}
\begin{figure}[h]
    \centering
    \includegraphics[width=0.49\textwidth]{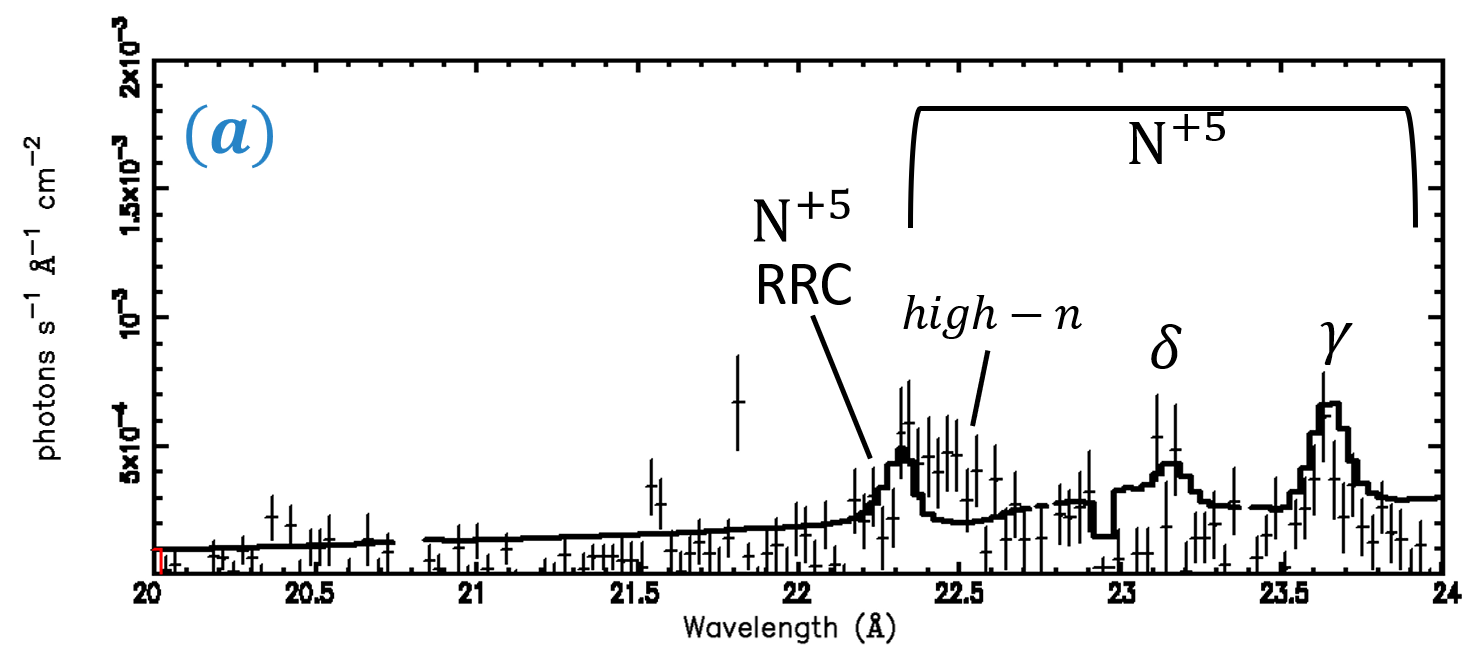}
    \includegraphics[width=0.49\textwidth]{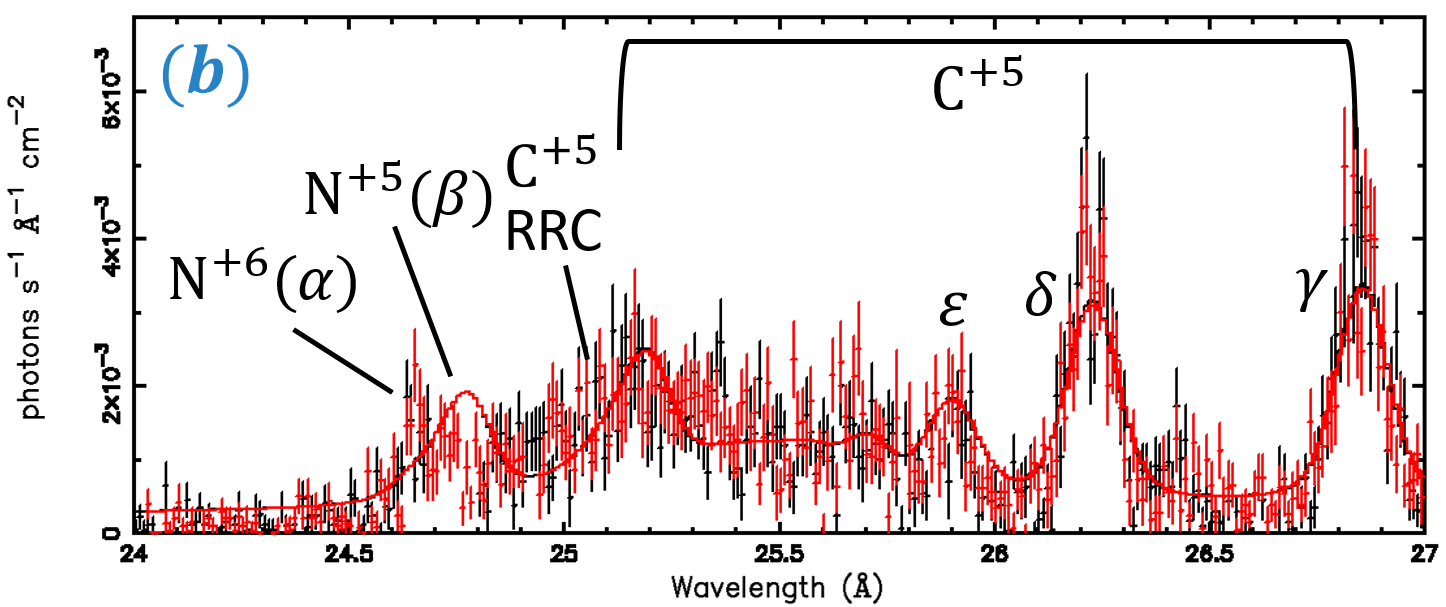}
    \includegraphics[width=0.49\textwidth]{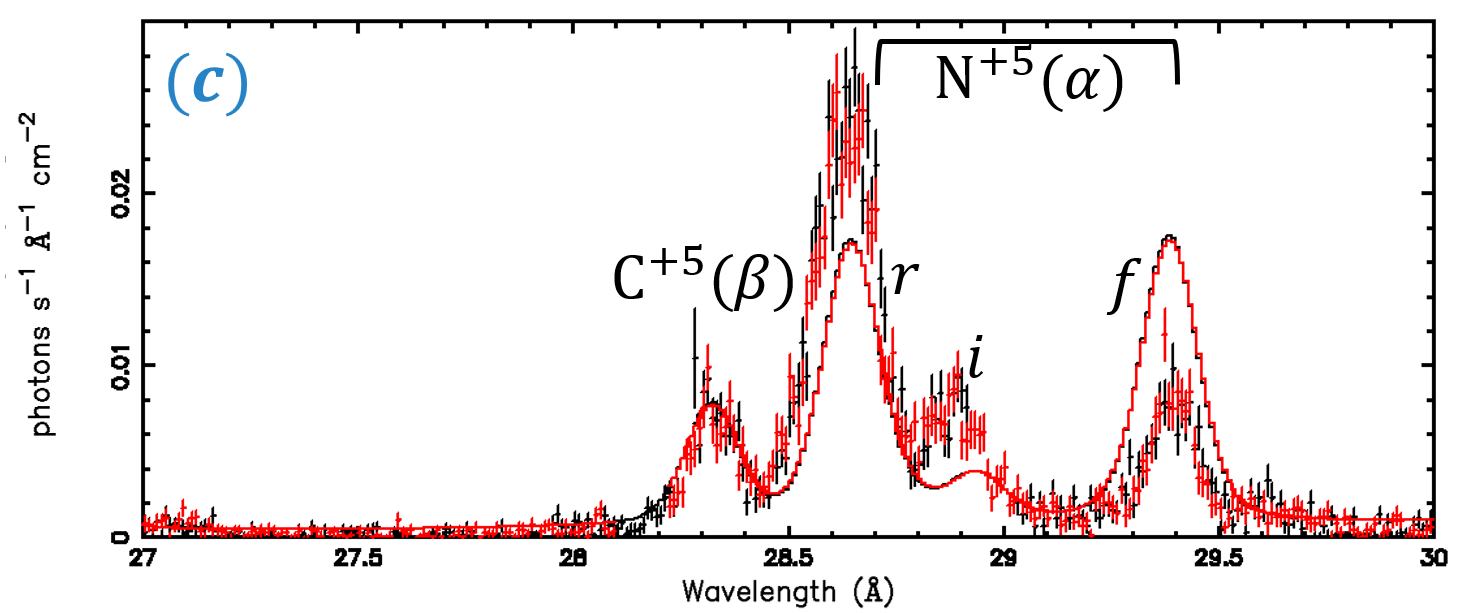}
    \includegraphics[width=0.49\textwidth]{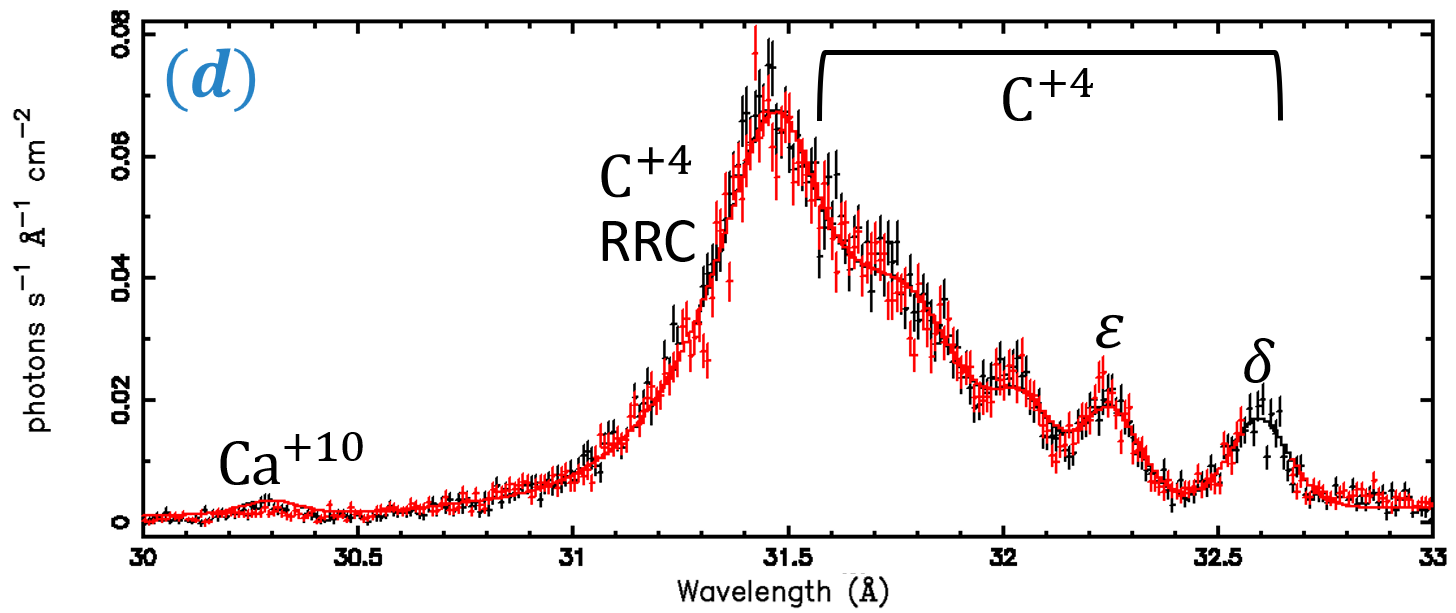}
    \includegraphics[width=0.49\textwidth]{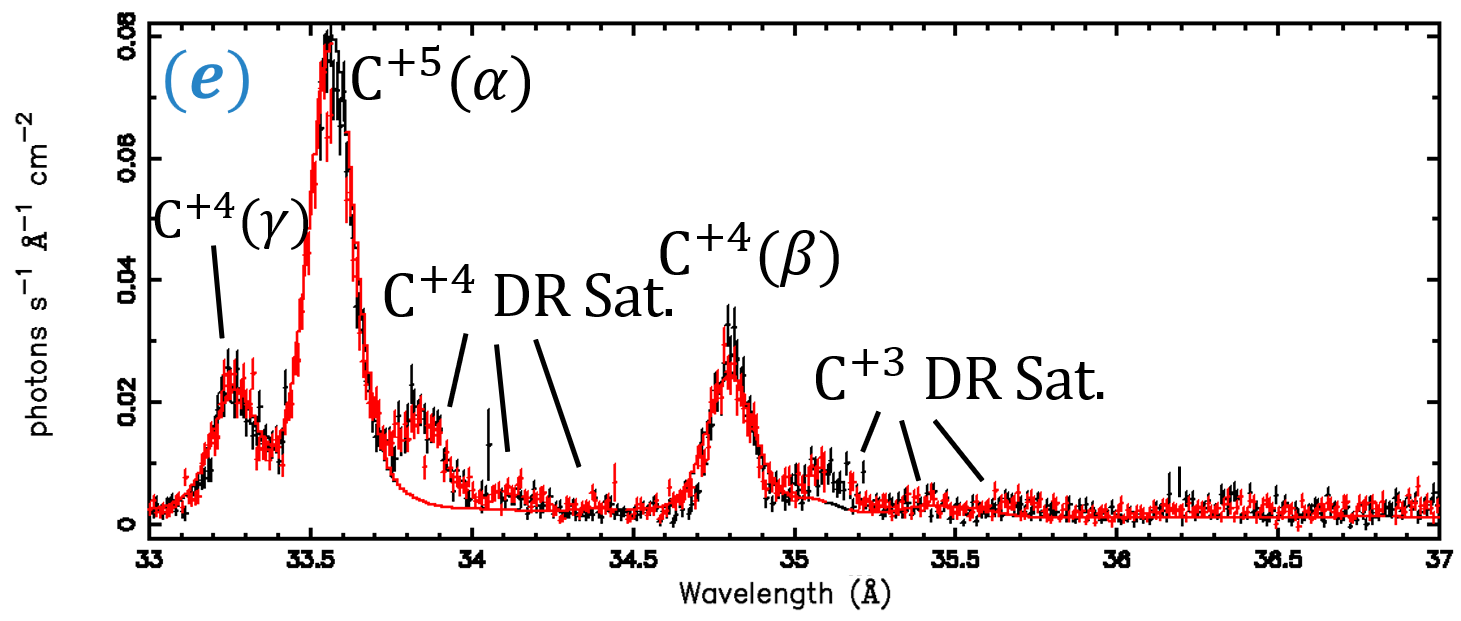}

  \caption{
    The full spectrum of YZ Ret observed by \xmm\ \rgs\ 1 (black) and \rgs\ 2 (red) in the range of 20-37\AA. The spectrum shows emission lines of several ions such as C$^{+4}$, C$^{+5}$, N$^{+5}$, N$^{+6}$, and Ca$^{+10}$. The solid line represents the fitted model of a velocity broadened collisionally ionized plasma (\textit{bvapec}) with 3 colder RRC components for C$^{+4}$, C$^{+5}$ and N$^{+5}$ (\textit{redge}), along with a series of gaussians to correct the amplitudes of the C$^{+4}$ and C$^{+5}$ high-$n$ transitions (\textit{agauss}). Note that the y-axis scale of the different panels is not the same.}
  \label{fig:full}
\end{figure}

We zoom into the \xmm\ \rgs\ spectrum (Fig.\,\ref{fig:yzret}) with five panels in Fig.\,\ref{fig:full}. The best-fitted model (solid line in the figure) 
is a low-density, collisionally-ionized plasma model with varying abundances. The C abundance is set the unity. This is complemented by the RRCs and high-$n$ line series of Sec.\,\ref{subsec:RRCdiagnostics}. In Xspec \citep{Arnaud1996}, %the model is \textit{tbabs}*(\textit{bvapec}+\textit{redge}(C$^{+4}$)+\textit{agauss series}(C$^{+4}$)+\textit{redge}(C$^{+5}$)+\textit{agauss series}(C$^{+5}$)+\textit{redge}(N$^{+5}$)). 
the \textit{bvapec} component \citep{Smith2001} represents an outflowing ($v=-1504\pm12$\,km\,s$^{-1}$) collisionally ionized plasma at $kT_e=73\pm1$\,eV, with lines kinematically broadened by $\sigma_v=520\pm4$\, km/s. Three \textit{redge} components represent the colder $\sim2$\,eV RRCs of C$^{+4}$, C$^{+5}$, and N$^{+5}$. The series of \textit{agauss} Gaussian lines represent the high-$n$ lines converging to the C$^{+4}$ and C$^{+5}$ RRCs (Sec.\ref{subsec:RRCdiagnostics}). 
The model also includes a Galactic absorption \textit{tbabs} component with $N_H=1.24\times10^{20}$\,cm$^{-2}$ \citep{Kalberla2005}. 
%When adding these spectral components, the overall fit improves drastically.
%and becomes statistically acceptable with 
%\textcolor{red}{The fit achieves C$_{stat}=10493$\,(2853 d.o.f). Sharon, this seems high, is it correct? maybe not mention?}

\begin{figure}[h]
    \centering
    \includegraphics[width=0.8\textwidth]{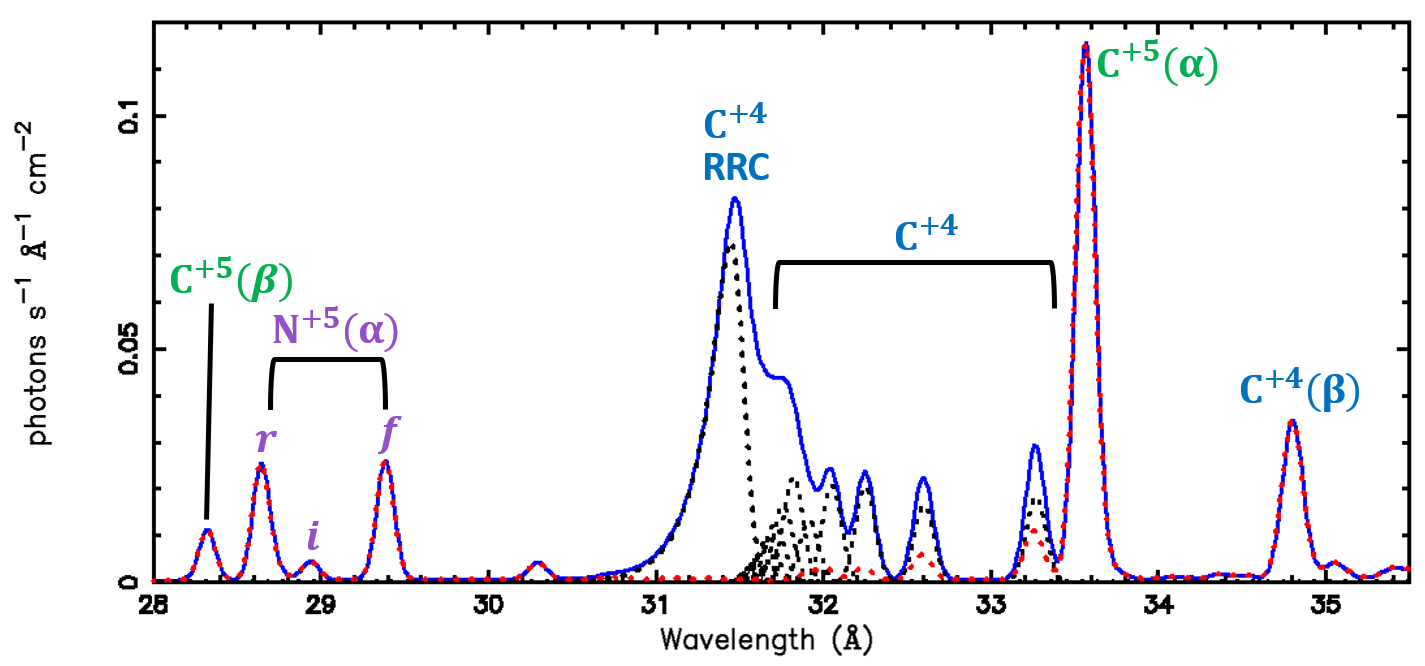}

  \caption{Theoretical model for the \rgs\ spectral range of 28-35.5\,\AA. The hot component (\textit{bvapec}) is plotted in (dotted) red. The RRC and the Rydberg series, which arise from the mixing with the cold medium,  are plotted in (dotted) black, and the total in (solid) blue. 
  }
  \label{fig:fullmodel}
\end{figure}

The hot plasma model (red dotted line in Fig.\,\ref{fig:fullmodel})   tightly constrains the outflow velocity and plasma temperature, yields the abundances of N, Ca, and provides an upper limit for the O abundance (see Table\,\ref{tb:abund}).
However, it cannot produce the RRCs, the high-order line series (black dotted lines in Fig.\,\ref{fig:fullmodel}), or the DR satellite lines. 
%Sharon, I commented out this paragraph, as I think all the information was just stated above.
%The {\it bvapec} component tightly constrains the overall temperature of the plasma based on the observed ions, and its outflow velocity based on the blue-shifts of the lines. 
%However, the brightest features in the spectrum discussed in the previous sections, most conspicuously the RRCs, indicate there is another component, which is colder (few eV) and dense ($\sim 10^{11}$\,cm$^{-3}$).
%
More specifically, the observed intensities of the Lyman and Helium line series of C$^{+5}$ and C$^{+4}$ are by far stronger in the spectrum than their hot-plasma values, starting already from $n=4$ ($\gamma$) and up to much higher and unresolved $n$ values. 
These series can be seen in panels (b) and (d) of Fig.\ref{fig:full}, where the hot plasma model has been appreciably enhanced with a series of Gaussians as in Sec.\,\ref{subsec:RRCdiagnostics}.
% that indicate a cold and dense component. 
%Another important spectral feature is 
The DR satellites of C$^{+4}$ and C$^{+3}$ are observed, respectively, between 33.8-34.4\,\AA\ and between 35-35.8\,\AA\ (panel (e) of Fig.\ref{fig:full}), and are also not nearly as strong in the model. 
These lines are calculated separately in Sec.\,\ref{sec:dr} (Fig.\,\ref{fig:drsat}). 
%This is consistent considering we observe a high flux for the RRC and high-$n$ series of C$^{+4}$/C$^{+5}$. 
Finally, the relative intensities of the He$\alpha$ N$^{+5}$ lines observed between 28.6-29.4\,\AA\  (panel (c) of Fig.\ref{fig:full}) and of He$\beta$ at 24.76\,\AA\ (panel (b) of Fig.\ref{fig:full}) are inconsistent with the model intensities, due to its low density approximation (Sec.\ref{subsec:Ndiagnostics}). 

\subsection{\chletg}
\begin{figure}
    \centering
    \includegraphics[width=0.8\textwidth]{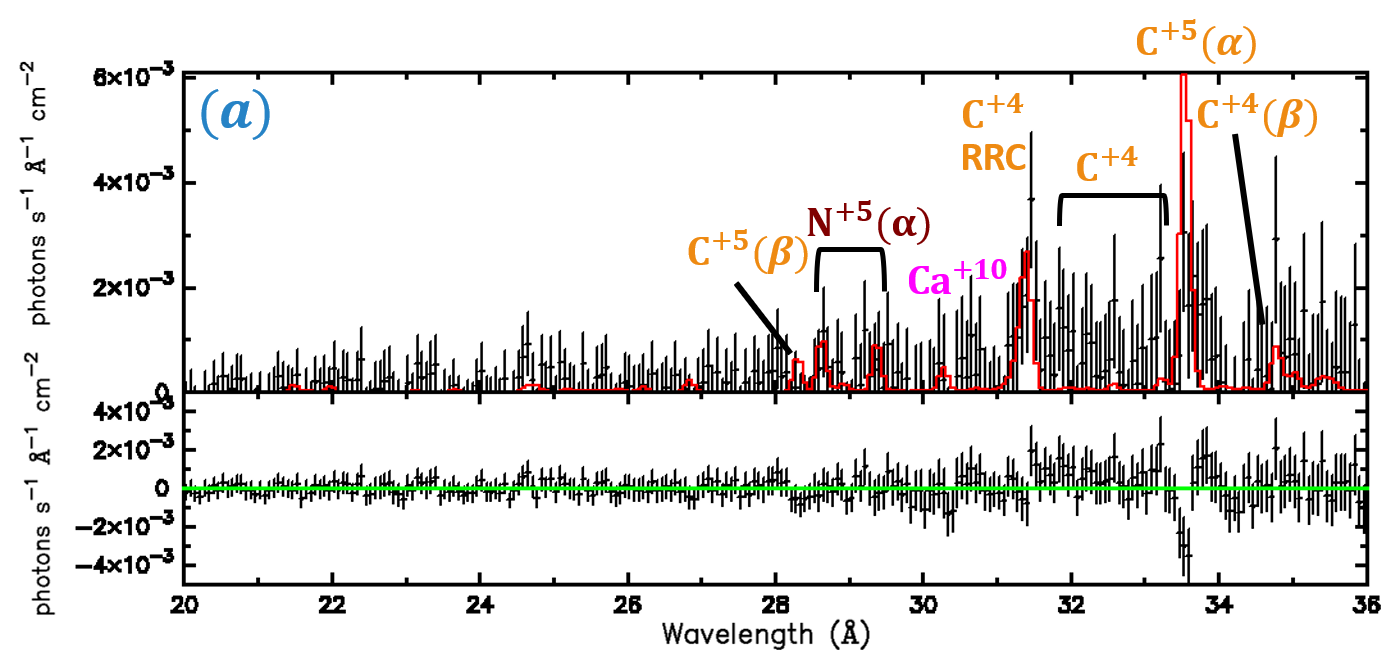}
    \includegraphics[width=0.8\textwidth]{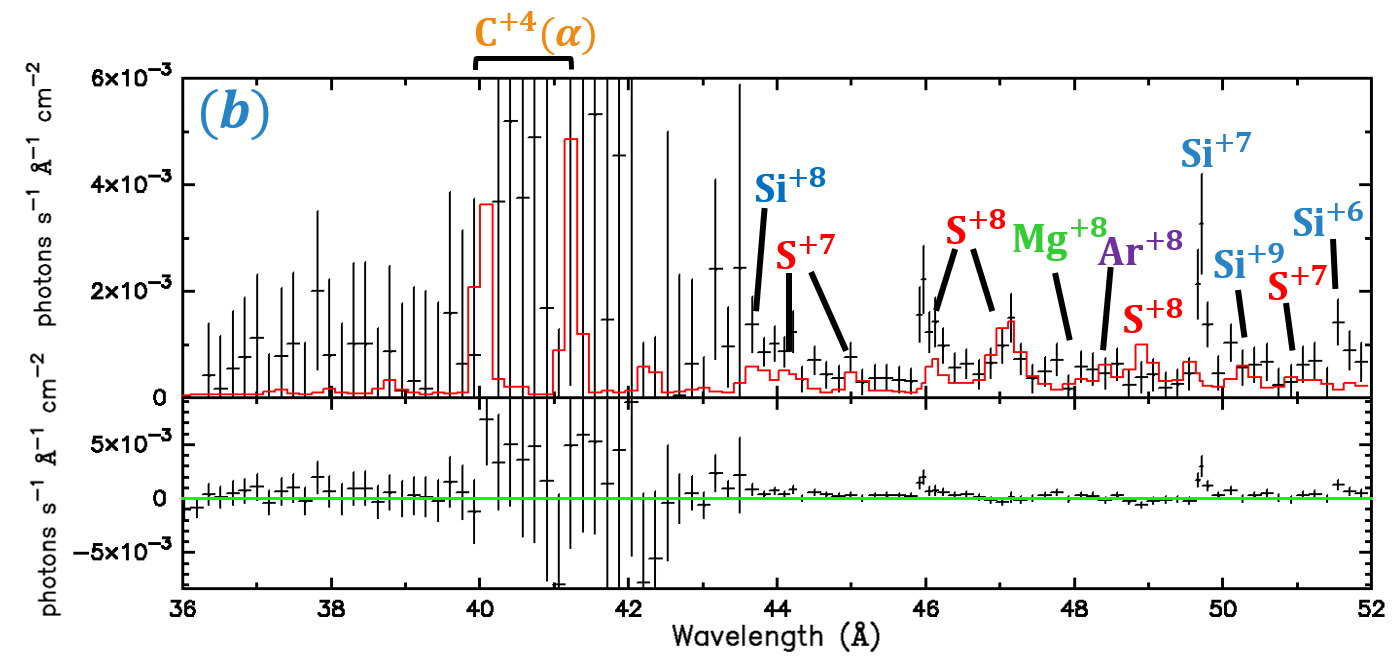}
    \includegraphics[width=0.78\textwidth]{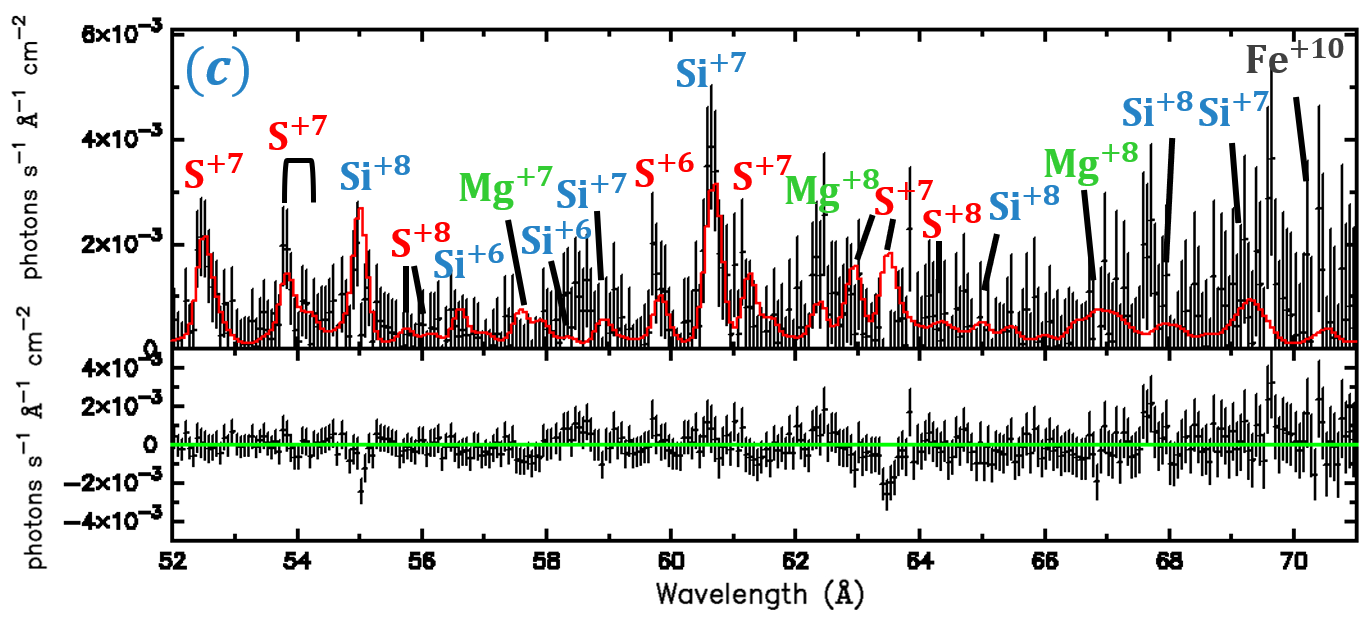}
 \caption{The full spectrum of YZ Ret observed with Chandra and the HRC-I+\letg\ instrumental setup. The spectrum is presented with the fitted model, the residuals (lower panels) and the identified lines. The cold temperature C$^{+4}$ RRC is still visible at 31.5\,\AA. In panel (b), where the effective area around the carbon edge of the detector is particularly low, the data are rebinned for visual clarity.} 
 \label{fig:chanfull}
 \end{figure}

\yzret\ was observed with the \chandra\ \letg\ (Fig.\ref{fig:chanfull}), 38 days after the \rgs\ observation. The \letg\ extends the \rgs\ waveband (up to 37\AA ) to 71\AA , but the source at this stage was much fainter. We fitted the \letg\ spectrum with a similar model to that of the \rgs. %with a goodness of fit C$_{stat}=$4234\,(3199 d.o.f), \textcolor{red}{which is better than for the \rgs\ due to the poorer S/N.} 
As in the \rgs\ spectrum, the prominent features in the \letg\ spectrum are the C$^{+4}$ RRC and the thermal line emission.
The C$^{+5}$ RRC and high-$n$ lines are present, but too faint to be fitted.

The \textit{bvapec} component represents an outflowing ($v=-1468\pm90$\,km\,s$^{-1}$) collisionally-ionized plasma at $kT_e=65\pm5$\,eV. The kinematic broadening can not be constrained and fixed to the \rgs\ best-fit value $\sigma_v=520$\, km/s. The RRC of C$^{+4}$ indicates a cold temperature of $kT_e<2.1$\,eV. The softer spectrum of \letg\ allows us to constrain the abundances of Mg, Si, S, Ar, and Fe that were not available with the \rgs .
%It represents the collisionally ionized plasma with temperature of kT=$86\pm4$\,eV and a redshift of $z=-(5.6\pm0.3)\times10^{-3}$, which corresponds to a velocity of $v=-1678\pm90$\,km/s towards the observer. The hot plasma temperature and velocity are slightly higher than the measured values of the \rgs\ spectrum. The C$^{+4}$ RRC is still visible, with temperature of kT$_{e,C^{+4}}<2.1$\,eV. The fit is done with the same broadening and galactic absorption. 
The measured parameters from \rgs\ and \letg\ fitted spectra are compared in Tables\,\ref{tb:rrc}, \ref{tb:abund}.

\begin{deluxetable}{lllllll}[h]
\tabletypesize{\footnotesize}
\tablewidth{0pt}
\label{tb:rrc}
\tablecaption{Best-fit spectral parameters obtained from \xmm\ \rgs\ and \chandra\ \letg .}
\tablehead{Observation&Component&Temperature\,[eV]&Velocity\,[km/s]&$\sigma_{\nu}$\,[km/s]&Flux\,[$10^{-3}$\,ph\,s$^{-1}$ cm$^{-2}$]&Volume\,[$10^{35}$\,cm$^{3}$]
}
\startdata
{\xmm\:\rgs} & $\textit{bvapec}$ &$73\pm1$ &$-1504\pm12$&$520\pm4$&$(184\pm2)^{\,(a)}$&$(6\pm3)$\\
{ } & $C^{+4} RRC$ & $1.96\pm0.04$ &"&"&$(24.2\pm0.4)$&$(8\pm4)$\\
{ } & $C^{+5} RRC$ & $1.9\pm0.2$ &"&"&$(0.5\pm0.05)$&-\\
{ } & $N^{+5} RRC$ & $<1.5$ &"&"&$(0.05\pm0.02)$&-\\
{\chandra\:\letg } & $\textit{bvapec}$ & $65\pm5$ &$-1468\pm90$&"&$(4.4\pm0.5)^{\,(a)}$&$(50\pm20)$\\
{} & $C^{+4} RRC$ & $<2.1$ &"&"&$(0.9\pm0.3)$&-\\
\enddata
\tablecomments{($a$) Measured in the range of 0.3-0.7 keV}
\end{deluxetable}

\begin{deluxetable}{c|llll}[h]
\tabletypesize{\footnotesize}
\tablewidth{0pt}
\label{tb:abund}
\tablecaption{Elemental abundances of the hot gas relative to solar \citep{Wilms2000} as measured by the \textit{bvapec} model. Unlisted elements are not observed and were taken at solar values.}
\tablehead{Element&\xmm\ \rgs\ & & Chandra \letg\ &
}
\startdata
{C}&1.00&&1.00&\\
{N} &$0.81\pm0.03$&&$0.7\pm0.4$&\\
{O}&$<5\times10^{-4}$&&$<0.2$&\\
{Mg}&-$^{(a)}$&&$1.5\pm0.6$&\\
{Si}&-$^{(a)}$&&$1.1\pm0.4$&\\
{S} &-$^{(a)}$&&$1.7\pm0.4$&\\
{Ar} &-$^{(a)}$&&$1.5\pm0.4$&\\
{Ca} &$1.2\pm0.1$&&$<6.6$&\\
{Fe} &-&&$4.0\pm3.3$&\\
\enddata
\tablecomments{($a$) Set to the values obtained by the Chandra fitted spectrum.}
\end{deluxetable}

\section{Discussion} 
\label{sec:discussion}
The key diagnostic is the high density $n_e=10^{11}$\,cm$^{-3}$, 
which implies extremely short time scales. 
For a C$^{+5}$ recombination rate coefficient of $\alpha^{RR}\sim3\times10^{-13}$\,cm$^{3}$\,s$^{-1}$, the recombination (and ionization) time is $\sim30$\,s.
For a solar abundance plasma at $kT=70$\,eV, the cooling function is $\Lambda\simeq 10^{-22}$\,erg\,cm$^3$s$^{-1}$ \citep{Sutherland1993}, and thus the radiative cooling time is $\tau = kT/n_e\Lambda \simeq 10$\,s.
The fact that the X-ray source persists on much longer time scales implies continuous heating, which can be attributed to a reverse shock in the nova ejecta.
Only the most recently shocked gas remains hot, resulting in a %an expanding 
thin X-ray emitting shell, consistent with the %X-ray emitting shell in 
theory of \citet{Hachisu2023}.
The reverse shock velocity can remain constant for a long time as it moves through the ejecta \citep{Hachisu2023}, as also expected in supernova remnants \citep[][Fig.\,4 therein]{Micelotta2016}.

If we ascribe a width $\delta r \approx v \tau \approx 10^9$\,cm to the shell during the \rgs\ epoch, the thin shell geometry implies $V=\Omega r^2 \delta r$, where $\Omega$ is the opening angle of the flow in sr. 
For $V_{\rm hot}=6\times 10^{35}$\,cm$^3$; $\Omega r^2 = 6\times 10^{26}$\,cm$^2$.
From here, the mass outflow rate in the shell is $\dot M=\Omega r^2 n v \mu m_p$. Plugging in all the above estimated parameters, we obtain $\dot M=3 \times 10^{-4}\,M_\odot$\,yr$^{-1}$.
Since in this approximation, $\delta r
\propto \tau \propto 1/n_e$, as the outflow expands and the density decreases, $\delta r$ increases.
In Section\,\ref{sec:dr}, we estimated the length scale of mixing between the hot and cold gas to be even smaller than $10^9$\,cm. 
%but two orders of magnitude higher than our other estimations (Sections \ref{sec:dr}, \ref{sec:2episodes}, and \ref{subsec:coolcomp}).

%with a compression ratio of 4, the shocked plasma velocity $v_p$ which we measure at 1500\,km\,s$^{-1}$ is related to the unshocked ejecta velocity ($v_{ej}$ and the shock velocity as follows: $v_{ej}=v_s = 4(v_p-v_s)$, or $3v_s=4v_p-v_{ej}$. Interesingly, for the escape velocity from a WD v_{ej} = 6000km/s, we get exactly v_s=0 (in the observer frame), which is what Micellota2016 are saying. The temperature would be detecrmined by (v_ej-v_s)^2=(6000km/s)^2 ~ 100keV.
%\delta r = (v_p - v_s) ~ tau*v_p.

\subsection{Shock heating and cooling}
\label{sec:shocks}

Direct evidence for the shocked plasma comes from the line widths. The measured $\sigma _v = 500$\,km\,s$^{-1}$ corresponds to a post-shock C ion ($M_C=12m_p$) temperature of $kT=12m_p\sigma _v^2 = 30$\,keV.
The reason we do not observe such high electron temperatures and fully charged ions is a combination of the 
long electron-ion interaction times, and the efficient electron cooling.
At $10^{11}$\,cm$^{-3}$, electrons would need $\sim 200$\,s to reach the ion temperature \citep[Fig.\,4.5 in][]{Vink2020}, while their cooling times are tens of seconds.

Assuming 30\,keV is the shock temperature $kT_s$, the relative velocity $v_s$ between the shock and the unshocked ejecta (neglecting radiation losses) is given by 
$kT_s=(3/16)m_pv_s^2$,
or $v_s=4000$\,km\,s$^{-1}$. 
The reverse shock is often much slower in the observed frame \citep{Micelotta2016}. Thus, $v_s$ is approximately the velocity of the unshocked ejecta. 
As expected, it is higher than 1500\,km\,s$^{-1}$, the velocity of the plasma after it is shocked.
The high velocity of $4000$\,km\,s$^{-1}$ suggests the ejecta escaped from close to the photosphere of the WD, yet is shocked further out.

\subsection{Comparing the two epochs, evidence for a standing shock?}
\label{sec:epoch}
What can we learn from comparing the \rgs\ (day 77) and \letg\ (day 115) spectra? Both are fitted well with a 70\,eV plasma model ({\it bvapec}). The lines in both are blueshifted by $\sim -1500$\,km\,s$^{-1}$, and broadened by $\sigma _v \sim 500$\,km\,s$^{-1}$.
Both spectra feature narrow ($\sim 2$\,eV) RRCs representing a cold plasma component.
However, the X-ray flux on day 115 is about 40 times lower (c.f. Fig.\,\ref{fig:swiftlc}), which is approximately true separately for both the hot and recombining components. Table.\,\ref{tb:rrc} summarizes the important fitted parameters.

If the X-ray source is moving out at 1500\,km\,s$^{-1}$ over these 38\,days, it traveled a distance of $5\times 10^{14}$\,cm.
Since the X-rays started to rise $\approx 20$\,days earlier, %\citep[c.f.][]{Hachisu2023}, 
by day 77 it could have already reached $\sim 2.5\times 10^{14}$\,cm.
%These distances are two orders of magnitude higher than the distance inferred if the plasma was photo-ionized (Sec.\,\ref{subsec:Ndiagnostics}), which rules  out photoionization.
Interestingly, Nova Delphini 2013 was also observed with gratings late (days 87 and 114 there), featuring blueshifted C and N lines, but in absorption with the SSS as a back-lighter \citep{Milla2023}. 
The fast velocities at these late times led those authors as well to explain the absorption by a thin shell ejected early, but now $10^{15}$\,cm away from the center.
The X-ray source in \yzret\ seems to be in a similar situation, only the SSS is obscured.

%We can now use these distances to infer a {\it lower limit} to the plasma emitting volume in the \letg\ epoch, where we have no good density measurement.
%The close redshift values measured in the \rgs\ and LETG spectra implies the plasma is traveling towards the observer at almost constant velocity. By that assumption, we compute the distance the plasma travelled until the \rgs\ observation to be $d_{travel}^{(77d)}=(1.00\pm0.01)\times10^{15}$\,cm. The distance the emitting plasma travelled in the 38 days between the \rgs\ and LETG observations is (taking the average velocity measured in the two observations) $d_{travel}^{(77-115d)}=(5.2\pm0.2)\times10^{14}$\,cm.
%We can infer the electron density at the Chandra LETG observation by using the measured flux, travelled distance, and the volume computations. The inferred density is dependent on the plasma geometry. 
%If the plasma is an expanding ball of gas, thus $V\propto r^3$, we obtain a radius of $r=(8.4\pm1.4)\times10^{11}$\,cm using $V_{bvapec}$ at the time of the \rgs\ spectrum. At the LETG observation, the gas expands to a volume of $V=(1.4\pm0.1)\times10^{44}$\,cm$^3$. By using The LETG measured flux (Table.\,\ref{tb:rrc}) and Eq.\,\ref{eq:emission_meas_vapec}, we infer an electron density of $n_e=(1.7\pm0.2)\times10^6$\,cm$^{-3}$.
 Returning to the conical outflow geometry, and using a distance at the \rgs\ epoch of $r=2.5\times 10^{14}$\,cm, and $V_{\rm hot}=\Omega r^2 \delta r = (6\pm3)\times 10^{35}$\,cm$^3$, we get $\Omega \delta r=(8\pm3)\times10^{5}$\,cm.
 If $\delta r \approx 10^9$\,cm, it implies a minuscule $\Omega$ of $8\times 10^{-4}$\,sr, namely small and dense blobs of ejecta.
 If on the other hand $\delta r \approx 10^7$\,cm (Sec.\,\ref{sec:dr}), then  $\Omega \sim 8\times 10^{-2}$\,sr, or a few degrees. %needs to be very small.
%Since nova outflows are clumpy, the shell is likely wider by one over the volume filling factor.  
%$r=d_{travel}^{(77d)}$, we obtain a shell thickness of $\delta r=(5\pm2)\times10^{4}$\,cm
 %radius of $r=(7.745\pm1.9)\times10^{17}$\,cm (up to a constant of $\delta r^{-1/2}$).
 At the \letg\ epoch, the gas expanded by a factor of 3 to $r=7.5\times10^{14}$\,cm.
%d_{travel}^{(77d)}+d_{travel}^{(77-115d)}=(1.52\pm0.02)\times10^{15}$\,cm, 
Assuming $\Omega \delta r$ remained the same, given similar cooling times, the corresponding volume on day 115 increased by a factor of 9, $V_{\rm hot} \propto r^2 = (5\pm2)\times10^{36}$\,cm$^{3}$.
The day 115 \letg\ X-ray flux is a factor $\sim40$ lower than on day 77. Thus, from Eq.\,\ref{eq:emission_meas_vapec}, $n_e$ decreased by only a factor of 2, and so did the cooling time, justifying an approximately constant $\delta r$.
%we infer an electron density of $n_e=(1.6\pm0.4)\times10^{10}$\,cm$^{-3}$, approximately an order of magnitude below the measured density in the \rgs\ spectrum, but
%still higher than the models of \citet[][Fig.\,8 therein]{Hachisu2023}.

%\subsection{Standing shock} \label{sec:2episodes}
The continuous shock over 38 days in Sec.\,\ref{sec:epoch} suggests a large distance from the center, and poses a challenge with its %and consequently an e
extremely small opening angle $\Omega$.
However, a reverse shock does not need to expand so fast, or at all. 
In a standing shock, the ejecta flows from the center at several 1000\,km\,s$^{-1}$, collides with the shock front, and exits the shock region at 1500\,km\,s$^{-1}$ (see Sec.\,\ref{sec:shocks} for the velocity and temperature estimates), while the shock location remains constant in the observer's frame of reference.
Since the cooling time is short, we observe fresh (recently) shocked ejecta at each time.
The fact that the temperature and velocity broadening are similar after 38\,days, attests to the constant relative velocities of the ejecta and the shock over this long time scale. 
In this scenario, the X-ray flux variability in Fig.\,\ref{fig:swiftlc} can be ascribed to the varying mass outflow rate of the nova.
A factor of 40 fainter X-rays with \letg\ compared to the \rgs , thus indicates a similar factor reduction in the density, from 10$^{11}$\,cm$^{-3}$ to a few 10$^9$\,cm$^{-3}$.
%An alternative scenario is that \rgs\ and \letg\ caught two separate ejection episodes by the WD atmosphere.
For a standing shock unfortunately, we have no distance diagnostics.
In the thin shell geometry, %These could occur 
a location closer to the center implies a larger opening angle, since the measured volume is $V \propto \Omega r^2$.
If, say, $\Omega \sim 1$\,sr, with $\delta r \sim 10^9$\,cm, it would imply a distance of $r= {\rm few} \times 10^{13}$\,cm, and higher for smaller $\delta r$.
%The spectral similarity between the \rgs\ and \letg\ observations implies it is the same shock-heated plasma, which allowed us to derive the distances and volumes, as discussed in 
%Sec.\,\ref{sec:epoch}. 
%Assuming that these are two independent emission episodes can imply a different physical picture than described in our analysis so far. 

\subsection{The Cool Plasma Component} \label{subsec:coolcomp}
The outflow velocity and the volume estimates of the cold plasma (2\,eV from the RRCs) are almost identical to that of the hot plasma, which suggests the two components are physically connected.
This is not the first time narrow RRCs are observed in shocked X-ray sources. 
\citet{Schild04} found C$^{+4}$ and C$^{+5}$ RRCs in the \rgs\ spectrum of the colliding wind binary $\gamma ^2$ Vel, which indicate a temperature of 40kK (few eV) in a hot (keV) plasma. 
\citet{Nordon2009} analyzed the  C$^{+5}$ RRC in the \letg\ spectrum of the planetary nebula BD$+30^\circ3639$, and found a temperature of $kT_e=1.7\pm1.3$\,eV in a hot 160-260\,eV plasma. 
These RRCs were unambiguously detected in high-resolution grating spectra.
Fe$^{+24}$ and Fe$^{+25}$ RRCs were detected with the {\it Suzaku} CCDs in supernova remnants \citep{Ozawa2009, Yamaguchi2009} at much higher temperatures of $kT \simeq 1$\,keV.
All of these findings combined suggest that narrow RRCs are a common feature of shocked plasmas.

The physical origin of these RRCs and the source of cold electrons are debated. 
While adiabatic cooling and ionization freezing has been proposed \citep{Schild04, Yamaguchi2009}, we suspect there is genuine mixing of cold electrons with the highly-charged ions.  
Mixing with cold gas can occur either at the reverse-shock front between shocked and unshocked wind, or at the contact discontinuity, at the interface between shocked ejecta and the dense, much cooler, shocked ISM gas. 
Computing the penetration depth of 2\,eV electrons into the hot gas region, by taking the Larmor radius in a $1\mu G$ magnetic field, one gets $R_L\simeq5\times10^6$\,cm, which is comparable to the hot thin shell size estimated in Sec.\,\ref{sec:dr}.
Neutral atoms from the cold phase of course penetrate the hot gas smoothly, with no magnetic obstacle.
These neutrals are discussed in Sec.\,\ref{sec:cx}.
%Larmor radius scales as 
%sqrt(mass)*sqrt(kT) 
%so hot ions will penetrate  sqrt(12*1846)*sqrt(100) more ~1500 deeper 
Another possible scenario is runaway radiative cooling. The cooling function of astrophysical plasmas peaks sharply at $\Lambda \approx 10^{21}$\,erg\,cm$^3$s$^{-1}$ between $4.2\,{\rm K} <\log T < 5.7$\,K \citep[e.g.,][]{Sutherland1993}.
This implies that gas cooling below $\sim 50$\,eV would rapidly cool down to 1-2\,eV. At the density of $10^{11}$\,cm$^{-3}$ this cooling time is $\tau < 1$\,s. This could explain why we observe narrow RRCs, namely no gas whatsoever at temperatures above 2\,eV.

The X-ray source in \yzret\ is exceptional in its high density, small volume, and prominent RRCs. We suspect that under these conditions, the highly charged ions are completely immersed in the dense cold gas, the source of which could be one of the above.  Another source of cold material could be an accretion disk.
\citet{Ness2013} suggested that X-ray emission line spectra observed with no SSS after nova eruptions originate in high-inclination systems, where the SSS is obscured by optically thick material, formed in a disk around it.
This disk can explain the lack of an observed SSS in \yzret, and provide a cold dense medium that the shocked gas runs into to produce the observed narrow RRCs and the CX. However, since the RRCs and high-$n$ lines are blueshifted, the wind would need to carry cool disk material with it without slowing down.
%The emission lines we observe would then come from the interaction between the WD winds and the obscuring disk.
%Although the obscuring disk is totally consistent with our observations, the separate ejection scenario is not free of difficulties either. First, the \xrt\ light curve (Fig.\ref{fig:swiftlc}) shows continuous emission that does not resemble two separate events. Second, 
The geometry and inclination of the system would need to be fine tuned to completely hide the WD, but allow us to still observe the approaching wind interacting with the edge of the disk. 

\section{Conclusions} \label{sec:conc}
We observed nova \yzret\ 2020 77 and 115 days after the eruption with \xmm\ and \chandra\ gratings, respectively.
The X-ray flux decreased by a factor of $\sim40$ between epochs.
Both X-ray spectra are characterized by emission lines with no sign of thermal emission from the SSS. 
In the following we summarize our conclusions from the analysis of the two spectra:

\begin{enumerate}

    \item Three totally different density diagnostics, namely using $f/i = 1.1\pm0.1$ in He-like N (Sec.\,\ref{subsec:Ndiagnostics}), $n_{\rm max}=35^{+10}_{-5}$ in the 1s$nl$ Rydberg series of He-like C (Sec.\,\ref{subsec:RRCdiagnostics}), and 1s$n'l'$ satellite line ratios of He-like C (Sec.\,\ref{sec:dr}), all point to $n_e \simeq 10^{11}$\,cm$^{-3}$.

  \item This key density diagnostic indicates radiative cooling and recombination times of the order of tens of seconds, which must imply ongoing heating, ascribed to a thin shock-heated layer of $10^{7}-10^{9}$\,cm in the nova ejecta (Sections \ref{sec:dr}, \ref{sec:discussion}).

  \item The emission lines originate in collisionally ionized plasma of $kT_e\simeq70$\,eV that was heated by the shock (Sec.\,\ref{sec:full}), but may not have had time to reach its maximal temperature or ionization degree due to the rapid radiative cooling. The $\sigma _v = 500$\,km\,s$^{-1}$ broadened lines suggest the ions may still be hot at $kT\sim30$\,keV.

  \item The narrow RRCs in the spectrum indicate the presence of cold 2\,eV electrons, recombining with the hot, highly-charged C and N ions (Sec.\,\ref{subsec:RRCdiagnostics}). These cold electrons could have crossed into the shocked region via the contact discontinuity, or the shock front (Sec.\,\ref{subsec:coolcomp}).

  \item The DR satellite line observed at 33.87\AA\ (rest frame) indicates a close interaction between the hot (70\,eV) and cold (2\,eV) components via photo-pumping. The high photon flux that is needed to produce the observed satellite line ratios implies the two are mixed %, which is consistent with the thin $10^7$\,cm shock-heated layer 
  (Sec.\,\ref{sec:dr}).

  \item The high-$n$
Rydberg series of C$^{+4}$ suggests highly charged ions are undergoing CX with neutral atoms, furthering the evidence for mixing of shocked and cold gas in the nova.

\end{enumerate}
The consistently high outflow velocity of $-1500$\,km\,s$^{-1}$ over 38 days would imply a source size of $\sim 10^{15}$\,cm (Sections 4.5, 6.1), but an outflow with a disturbingly minuscule opening angle, consistent maybe with small dense blobs and a low volume filling factor. 
Conversely, invoking a standing reverse shock could imply a shorter distance, and thus %bring the outflow closer, e.g., to interact with the accretion disk, by requiring that the two epochs are not related, 
alleviate the small opening angle problem. %, but creates other issues (Sec.\,\ref{sec:2episodes}).

\bigskip
S.M. and E.B. were supported in part by a Center of Excellence of the Israel Science Foundation (grant No. 1937/19).
M.O. was supported by a Chandra-Smithsonian award for the \chletg\ observation.
This paper employs a list of Chandra datasets, obtained by the Chandra X-ray Observatory, contained in the Chandra Data Collection (CDC) 233~\dataset[doi:10.25574/cdc.233]{https://doi.org/10.25574/cdc.233}

%\appendix

%\section{Appendix information}


\bibliography{bibli}{}
\bibliographystyle{aasjournal}



\end{document}